\documentclass[prb,preprint,samsmath,amssymb,floatfix,superscriptaddress,nofootinbib,aps]{revtex4}
\usepackage{graphicx}
\usepackage{epstopdf}
\usepackage{amsmath}

\begin{document}
\draft
\title{Theoretical and numerical study of the phase diagram of patchy colloids: ordered and disordered patch arrangements}
\author{Emanuela Bianchi} 
\affiliation{ {Dipartimento di Fisica and
  INFM-CRS-SMC, Universit\`a di Roma {\em La Sapienza}, Piazzale A. Moro
  2, 00185 Roma, Italy} } 
\author{Piero Tartaglia} 
\affiliation{ {Dipartimento di Fisica and
  INFM-CRS-SMC, Universit\`a di Roma {\em La Sapienza}, Piazzale A. Moro
  2, 00185 Roma, Italy} } 
\author{ Emanuela Zaccarelli}  \affiliation{ {Dipartimento di Fisica and
  INFM-CRS-SOFT, Universit\`a di Roma {\em La Sapienza}, Piazzale A. Moro
  2, 00185 Roma, Italy} } 
\author{  Francesco Sciortino} \affiliation{ {Dipartimento di Fisica and
  INFM-CRS-SOFT, Universit\`a di Roma {\em La Sapienza}, Piazzale A. Moro  2, 00185 Roma, Italy} } 
 
  \begin{abstract}
We report theoretical and numerical evaluations of the
phase diagram for a model of patchy particles. 
Specifically we study hard-spheres whose surface is decorated by
a small number $f$ of identical  sites (``sticky spots'') interacting via a short-range square-well attraction.
We theoretically evaluate, solving the Wertheim theory,  the location of the critical point and the gas-liquid coexistence line for several values of $f$ and compare them to results of Gibbs and Grand Canonical Monte Carlo simulations.  We study both ordered and disordered arrangements of the sites on the hard-sphere surface and  confirm that patchiness has a strong effect on the phase diagram: the gas-liquid coexistence region in the temperature-density plane is significantly reduced as $f$ decreases. We also theoretically evaluate the locus of specific heat maxima and the percolation line. 
\end{abstract}

\maketitle
\section{INTRODUCTION}

Patchy particles are particles interacting via a limited number of directional  interactions.
The anisotropy of the interaction leads to collective behaviors different from those of simple liquids. Gelation~\cite{Trappe,Cipelletti, zaccajpcm,advances,genova}, gas-liquid phase separation~\cite{Zacca1,bian}, crystallization~\cite{Wilber-2006,doye,octahedral}  and clustering are strongly affected by patchiness~\cite{Glotz_Solomon_natmat,zhangglotzer,Zhang_03,simone,bastiaan}.
Recently, a new generation of colloidal particles with chemically or physically patterned surfaces has been designed and synthesized in the attempt to provide valence to colloids~\cite{Manoh_03,mohovald,Blaad06, Cho_05,Glotz_Solomon,Glotz_Solomon_natmat}.  This relevant synthesis effort aims to generate super atoms --- atoms at the nano and micro-scopic level ---  in order to reproduce and extend the  atomic and molecular behavior on larger length scale. It also offers the possibility to export the supra-molecular chemistry ideas~\cite{leibler,lehn1,lehn2} to new  colloidal materials, opening the new field of supra-particle colloidal physics.   Thus, a general effort to develop a deeper understanding of self-assembly  and to construct a more unified theoretical underpinning for this technologically and scientifically important field is crucial. The outcome of this effort may also have an impact in our understanding of the phase behavior of protein solutions, due to their intrinsic patchy character~\cite{Lomakin,Sear_99,KumarJCP07,McManus}.

Our recent work~\cite{bian} has shown that the Wertheim theory~\cite{Werth1,Werth2} describes rather well the critical properties of  particles decorated on their surface by a predefined number of  attractive sites. The Wertheim theory is a thermodynamic perturbation theory introduced to describe association under  the hypothesis  of  a single-bond per patch, which means  that an attractive site on a particle cannot bind simultaneously to two (or more) sites on another particle. The single-bond per patch condition can be naturally implemented in colloids
by choosing an appropriate small ratio between the range of the attractive patches and the particles size. The single-bonding condition results also from  DNA complementarity~\cite{dna,dnastarr}  as well as from complementary ``lock-and-key'' interactions associated to biological specificity~\cite{hiddessen1,hiddessen2}.  These types of interactions provide a versatile way of controlling inter-particle binding.
An extension of the Wertheim  thermodynamic perturbation theory to interpret and/or predict the behavior of a wide range of substances with potential industrial applications
is provided  by the statistical associating fluid theory (SAFT)~\cite{Chap1,Chap2} and by its developments~\cite{Saft-vr,Saft-vre}.

In previous works we have shown~\cite{Zacca1,bian,23} that for patchy colloidal particles with a small number of sticky sites the critical point of the gas-liquid phase separation moves towards small packing fraction ($\phi$) and temperature ($T$) with decreasing the number of patches.
According to this study, liquid phases of vanishing density can be generated once a small fraction of polyfunctional particles is added to a system of bifunctional ones. 
Indeed, the study of binary mixtures of patchy particles with different functionalities allows to explore also the range of non-integer valence down  to $2$.
 This means that with the new generation of non-spherical sticky colloids~\cite{Manoh_03,Cho_05}, it should be possible to realize "empty liquids"~\cite{bian}, i.e. states  with an arbitrarily small occupied packing fraction at temperature lower than the liquid-gas critical temperature.
The shift  with valence of the critical point, both in density and temperature, leads to substantial changes in phase behavior with branching: the reduction of the number  of bonded nearest neighbours is accompanied by an enlargement of the  region of stability of the liquid phase in the $(T,\phi)$ plane. This fact could favor the establishment,   at low $T$ and at small $\phi$,  of  homogeneous disordered materials, i.e. equilibrium  disordered  states  in which particles are interconnected in a persistent network.  At such low $T$,  the bond-lifetime will  become comparable to the experimental observation time. Under these conditions,   it should be possible to approach dynamical arrested  states continuously from equilibrium and to generate a state of matter  as close as possible to an ideal gel~\cite{genova,zaccajpcm}.

In this article we extend the preliminary study of Ref.~\cite{bian}
reporting  a Monte Carlo investigation of the $f$-dependence of the critical point location for a model with a disordered arrangement of patches. The present study
confirms the trend  discussed in Ref.~\cite{bian}  for the  corresponding ordered case.  
To evaluate the role of the valence on the coexistence region, we also numerically  investigate the shape of the gas-liquid binodal line for the ordered  case and compare it with theoretical predictions  based on the Wertheim theory. Finally, we analytically  calculate several equilibrium properties, such as the energy per particle, the specific heat, the extent of polymerization and the percolation line, to get insights on their $f$-dependence.

We find that the reduction of valence is accompanied by a significant  shift of the coexistence curve towards low temperature and density. The percolation line is always found to lie above the critical point, merging with the gas-liquid spinodal at low density $\rho$. The liquid state is thus  always characterized by an infinite spanning network. This confirms the possibility of observing, for large attraction strengths, dynamical arrested states driven by bonding (as opposed to packing) in single phase conditions, i.e. homogenous arrested states at low density.

We also provide in the Appendix A a physical insight of the Wertheim theory by showing that the theoretical  expression for the free energy in Ref.~\cite{Werth1,Werth2} is formally equivalent to the free energy of a system  of non-interacting clusters distributed according to  the Flory-Stockmayer cluster size distribution~\cite{flory}.

\section{The model}\label{sec:model}

We focus on a system of particles modeled as hard-spheres of diameter $\sigma$, whose surface is decorated by $f$  bonding sites at fixed locations. Sites on different particles interact via a square-well potential. The  interaction $V({\bf 1,2})$  between particles {\bf 1} and  {\bf 2} is \begin{equation}
V({\bf 1,2})=V_{HS}({\bf r_{12}})+\sum_{i=1}^f\sum_{j=1}^f V_{SW}({\mathbf r}^{_{ij}}_{_{12}})
\end{equation}
where  the individual sites are denoted by $i$ and $j$,  $V_{HS}$ is the hard-sphere potential, $V_{SW}(x)$ is a square-well interaction (of depth $-u_0$ for $x \leq  \delta$, 0 otherwise) and  ${\bf r_{12}}$ and ${\mathbf r}^{_{ij}}_{_{12}}$ are respectively the vectors joining the particle-particle and the site-site (on different particles) centers. Geometric considerations for a  three touching spheres configuration show that the choice $\delta=0.5(\sqrt{5-2\sqrt{3}}-1) \sigma  \approx 0.119 \sigma$ guarantees that each site is engaged at most in one bond. Hence, with this choice of $\delta$,  each particle can form only up to $f$ bonds.  We note that in this model bonding is properly defined: two particles are bonded when their pair interaction energy is -$u_0$.  Distances are measured in units of $\sigma$. Temperature is measured in units of the potential depth (i.e. Boltzmann constant $k_B=1$).

We study two different arrangements of the $f$ sites on the particles surface. In the first case sites are arranged in a regular structure (see Fig.~1 of Ref.~\cite{bian}). In the second case, the distribution of the sites is random and different for each particle. In this latter case, the only constraint on the site position is formulated on the basis of a minimum distance $d_{min}$ criterion between different sites on the same particle: the choice of $d_{min}$ aims to minimize the possibility of double bonding between the same pair of particles as well as the shading of a bonding site by the presence of a nearby bonded site. We choose $d_{min}=0.4$. 

\section{The theory}
The  first-order thermodynamic perturbation Wertheim  theory~\cite{Werth1,Werth2,Hansennew} provides an expression for the free energy of  particles with a number $f$ of attractive sticky sites on their surface, independently from the specific geometric arrangement of the sites.  The theory assumes that all sites have the same probability of forming bonds and that the correlation between adjacent sites is missing. 

The Helmholtz free energy of the system is written as  a sum of the hard-sphere reference free energy, $F^{HS}$, plus a  bond contribution, $F^{bond}$. The Helmholtz free energy due to bonding derives from a summation over certain classes of relevant graphs in the Mayer expansion~\cite{Hansennew}. In the sum, closed loops graphs are neglected.   The fundamental assumption of the Wertheim theory  is that the conditions of steric incompatibilities are satisfied:  (i) no site can be engaged in more than one bond and (ii) no pair of particles can be double bonded. These steric incompatibilities are satisfied in both our models thanks to  (i) the small $\delta$ chosen for the short-ranged square-well attraction and  to (ii)  the location of the sticky sites on the  hard-sphere particles surface.
In the  formulation of Ref.~\cite{Chap1}, the bond free energy density of a system of $f$-functional particles is
\begin{equation}\label{eq:f}
\frac{\beta F^{bond}}{V} =  \rho \ln (1-p_b)^f +\frac{1}{2}\rho f p_b
\end{equation}
where $\beta=1/k_BT$,  $\rho=N/V$ is the particle number density and $p_b$ is the bond probability. Since we assume equal reactivity for all sites, the bonding process can be seen as a chemical reaction between two unsaturated sites in equilibrium with a pair of  bonded sites. In this respect one can write
\begin{equation}\label{eq:pb}
 \frac{p_b}{(1-p_b)^2}= \rho e^{-\beta {\cal F}_b}
\end{equation}
where ${\cal F}_b$ is the site-site bond free-energy, i.e. the free energy difference between the bonded and the unbonded state.

The Wertheim theory predicts an expression for ${\cal F}_b$  in term of liquid state correlation functions and spherically averaged Mayer functions.  More precisely
\begin{equation}\label{eq:deltaf}
e^{-\beta {\cal F}_b} = f \Delta
\end{equation}
where $\Delta$  refers to a single site-site interaction (since all bonding sites are identical) and it is defined as
\begin{equation}\label{eqn:Deltabis}
\Delta= 4 \pi \displaystyle \int_{\sigma}^{\sigma+\delta} {g_{HS}(r_{12})\langle f(12)\rangle _{\omega_{1},\omega_{2}} r_{12}^{2} d r_{12}}.
\end{equation}
Here $g_{HS}(r_{12})$ is the reference hard-sphere fluid pair correlation function, the site-site Mayer function is $f(12)=\exp[-V_{SW}({\mathbf r}^{_{ij}}_{_{12}})/k_{B}T]-1$, and   $\langle f(12)\rangle _{\omega_{1},\omega_{2}}$  represents an 
 angle-average over all orientations of particles 1 and 2 at fixed relative distance $r_{12}$. 
Since the Wertheim theory is insensitive to the location of the attractive sites, the number of interacting sites on each particle is encoded only  in the factor $f$ before $\Delta$ in Eq.~\ref{eq:deltaf}.
For a site-site  square-well interaction, the Mayer function can be calculated as~\cite{Werth5}
\begin{equation}
\langle f(12)\rangle _{\omega_{1},\omega_{2}} =  \left [
\exp(\beta u_0)-1 \right ] S(r) 
\end{equation}
where $S(r)$ is the fraction of solid angle available to bonding when two particles are located at relative center-to-center distance $r$ $(r\equiv r_{12})$, i.e.
\begin{equation}
S(r)=\frac{(\delta + \sigma -r)^2  (2 \delta - \sigma +r )}{ 6 \sigma^2 r}.
\end{equation}
The evaluation of $\Delta$ requires only an expression for
$g_{HS}(r)$  in the range where bonding occurs ($\sigma<r<\sigma+\delta$). 
We use the linear approximation~\cite{Nez_90} 
\begin{equation}
g_{HS}(r)= \frac{1-0.5 \phi}{(1- \phi)^3}-\frac{9}{2}\frac{\phi (1+\phi)}{(1- \phi)^3} \left[\frac{r-\sigma}{\sigma}\right]
\label{eq:ghsr}
\end{equation}
(where $\phi=\frac{\pi}{6} \sigma^3 \rho$), which provides the correct Carnahan-Starling~\cite  {CS_69} value at contact. This gives
\begin{eqnarray}
\label{eq:deltamodel}
\Delta= 
\frac{V_b (e^{\beta u_0} - 1)}{ (1-\phi )^3}  \times \\ \nonumber
\left[1-\frac{5}{2}\frac{\left(3\sigma^2+8 \delta\sigma +3 \delta ^2\right)  }{ \sigma(15\sigma+4 \delta ) }\phi -\frac{3}{2}\frac{\left(12 \delta \sigma +5 \delta
^2\right) }{\sigma (15\sigma +4 \delta ) }\phi ^2\right] 
\end{eqnarray}
where we have defined the spherically averaged bonding volume $V_b \equiv  4 \pi \int_\sigma^{\sigma+\delta} S(r) r^2 dr =\pi \delta^4 (15\sigma +4 \delta)/30\sigma^2$.
We note that the above expression of $\Delta$ simplifies in the low density limit. Indeed, when $\rho \rightarrow 0$, the hard-sphere pair correlation function tends to the ideal gas limit value $g_{HS}(r) \approx 1$.  In this limit $\Delta$ doesn't depend on $\rho$, i.e. $\Delta=V_b (e^{\beta u_0} - 1)$.
We note that bonding takes approximatively place when $\exp(\beta u_0) \gg 1$. Indeed bond formation arises from a balance between the energetic gain of forming a bond ($\Delta U_b=-u_0$) and an entropy loss ($\Delta S_b$), which is expressed in the theory as logarithm of the  ratio between $V_b$  and  the volume per bonding site,  $V/(f N)$~\cite{M2}. Since $V_b \ll  V/(fN)$, bonding becomes relevant when $\beta u_0 \gg 1$. In the following we will thus approximate ($e^{\beta u_0} - 1$) with $e^{\beta u_0}$ to simplify the theoretical expressions.

Once the free energy is known, it is possible to derive various equilibrium properties of the system through  thermodynamic relations. 
We find expressions for the energy per particle, the specific heat maxima, the extent of polymerization and the pressure of the system in terms of $p_b$, which is a function of $T$ and $\rho$ from Eq.~\ref{eq:pb}.
The potential energy per particle $E/N$ is given by
\begin{equation}\label{eqn:energy}
\frac{E}{N}= \frac{\partial (\frac{\beta F^{bond}}{N})}{\partial \beta}=-\frac{1}{2} f u_0  p_b
\end{equation}
i.e. it is exactly  the fraction of bond times $-u_0 f/2$. 
The constant volume specific heat $C_V$ can be calculated as
\begin{equation}\label{eq:cv}
C_V= \frac{\partial (\frac {E}{N})}{\partial T}= \frac{1}{2}f  \frac{u_0^2}{T^2}  \frac{p_b(1-p_b)}{1+p_b}. 
\end{equation}
At each $\rho$, the specific heat has a maximum (whose amplitude increases with $f$) at finite $T$, which defines a line of specific heat extrema in the $(T,\rho)$ plane. The $C_V^{max}$ line can be used as an estimate of the polymerization transition line~\cite{Greer88,Greer96,Greer02,Milchev98,douglas,Dudo_03}. 

In the characterization of the self-assembly of particles, experimentalists often consider a quantity called extent of polymerization $\Phi(t)$, which is normally measured by spectroscopy. $\Phi$ is defined as the fraction of particles bonded in clusters. This quantity plays the role of order parameter in the polymerization transition: it changes continuosly  form the value zero at high $T$, when all particles are in the monomeric state, to the value one at low $T$, when particles are bonded in clusters. This crossover becomes sharper and sharper on decreasing $\rho$.
Since the monomer density is simply obtained by the observation that all of the sticky spots on each particle must be unbonded, i.e. $\rho_{1} = \rho (1-p_b)^f$, the extent of polymerization is given by
\begin{equation}
\Phi= \frac{\rho -\rho_{1}}{\rho} = 1- (1-p_b)^f.
\end{equation}
As a function of $p_b$, the branched polymerization transition becomes sharper and sharper on  increasing the functionality of the system. 


The pressure $P$ of the system can be evaluated by deriving, respect to the volume, the Wertheim free energy, i.e. $\beta P=- (\partial \beta F/\partial V)_T$. The bonding contribution to $P$ is thus
\begin{equation}\label{eq:Pbond}
\frac{\beta P^{bond}}{\rho}=   \rho f \left[\frac{1}{2}-\frac{1}{1-p_b}\right]\frac{\partial p_b}{\partial \rho}.
\end{equation}
In the low $\rho$ limit ($g_{HS}(r) \approx 1$), it is possible to neglect the $\rho$ dependence of $\Delta$ and  $\beta P^{bond}/ \rho$ becomes equal to $ -\frac{1}{2} f p_b$. Appendix A provides a physical understanding of this expression.
The hard-sphere contribution to the pressure can be evaluated via the  Carnahan-Starling equation of state~\cite  {CS_69}
\begin{equation}\label{eq:betaPhs}
\frac{\beta P^{HS}}{\rho}= \frac{(1 + \phi + \phi^2 - \phi^3)}{(1 - \phi)^3}.
\end{equation}

From the resulting  $V$ and $T$ dependence of $P$, it is possible to evaluate 
 the liquid-gas coexistence region in the phase diagram, by solving 
 the following set of equations
 \begin{eqnarray}\label{eq:coex}
\nonumber
T_g = T_{l} \equiv T^* \\
P_g = P_{l} \equiv P^* \\
\nonumber
\int_{V_l}^{V_g}[P(V,T^*)-P^*] dV = 0,
\end{eqnarray}
where  $T_g, P_g, V_g$ and $ T_l, P_l, V_l$ are  respectively the temperature, the pressure and the volume of the two coexisting phases. The third equation corresponds to the Maxwell construction.

The main assumption of the Wertheim theory is that molecules (or particles) cluster in open structures without closed bond loops. The hypothesis of absence of closed bonding loops is also at the heart of the   Flory-Stockmayer 
theory, developed to model aggregation in chemical gelation. The Flory-Stockmayer theory~\cite{flory}  provides expressions for the number density of
clusters of $n$ particles, $\rho_n$, as a function of the bond probability (the extent of the reaction in the Flory-Stockmayer language). For functionality $f$ 
\begin{eqnarray}\label{eq:flory}
\rho_n &=& \rho   (1-p_b)^f \left[ p_b (1-p_b)^{f-2}\right]^{n-1}  \omega_n\\
\nonumber
\omega_n &=& \frac{f (f n-n)!}{(fn-2n+2)!n!}
\end{eqnarray}
where $\rho \equiv \sum_{n} n \rho_n= N/V$ is the total number density.
In Appendix A, we show that the Wertheim free-energy of Eq.~(\ref{eq:f}) is equivalent to the free energy of a system of non-interacting clusters distributed according to
Eq.~\ref{eq:flory}. Here we make use of the  Flory-Stockmayer theory for 
providing an expression, to be used in conjunction with the bond probability derived using the Wertheim theory, to evaluate the location in the $(T,\rho)$ plane of the percolation line. The  bond probability at percolation, $p_b^p$, is
\begin{equation}\label{eq:perc}
p_b^p=\frac{1}{f-1}.
\end{equation}

Fig.~\ref{fig:pdWtot} shows the resulting phase diagram evaluated according to the Wertheim theory for three different values of $f$. More specifically it shows 
the relative location of  the
 phase coexistence line, the percolation  and the maxima of specific heat line. According to the Wertheim theory, the
 coexistence region becomes wider on increasing $f$.  For the case
 $f=5$, at low $T$ the gas coexists with a liquid with number density
 $\rho \approx 0.8$, a value significantly smaller than the one commonly observed for particles interacting via spherical potentials. 
 The percolation line merges into the coexistence curve on the left of the
 critical point, confirming that a spanning cluster of bond is a pre-requisite for
 the development of a critical phenomena~\cite{coniglio}.  For the shown  $f$ values, the locus of $C_V^{max}$ is located  below the corresponding percolation line, in agreement also with recent findings for a spherical model with $f=4$~\cite{prossimo}.
However  in the limit where $f \rightarrow 2$, realized via a mixture of 
$f =2$ and $f=3$ particles with average functionality $2.055$~\cite{23}, 
 the percolation line lies below the $C_V^{max}$ line.
  The intersection of the $C_V^{max}$ line with the coexistence curve  progressively moves from the left to the right of the critical point on increasing $f$.  Already for $f=5$ the density at which  the $C_V^{max}$ line meets the
 coexistence line is more than twice the critical density.  While it is not
 reasonable to extend the Wertheim theory to large $f$ values,
 it is tempting to
 speculate that, on further increasing $f$, the intersection point will keep on 
 moving to larger densities so that in the spherical limit case ( with analogous range of interaction)  the entire  $C_V^{max}$ line lies  in a physically inaccessible region (due to packing-driven kinetic arrest).
  

\section{Monte Carlo Simulation}

We perform simulations of the first model discussed in Sec.~\ref{sec:model}
(in which the sticky spots location is regular) with the aim of evaluating the gas-liquid coexistence lines. We aim to provide a definitive proof that reducing valence generates a region of thermodynamic stability of the liquid phase down to vanishing temperatures in a wide range of densities.
Previous studies of the same models were indeed focused only on the location of the critical point~\cite{bian}.
We perform Gibbs Ensemble Monte Carlo simulations (GEMC) in order to evaluate the phase coexistence region of one component systems with functionality $f$.  
The GEMC method~\cite{GEMC} allows us to study  coexistence in the region where the gas-liquid free-energy barrier is sufficiently high to avoid crossing between the two phases.  We simulate for about 5 million MC steps, where a MC step is defined as $N_{\Delta}=10^5$ attempts to translate and  rotate a randomly chosen particle, $N_{N}= 10^3$ attempts to swap a particle between the gas and the liquid boxes and $N_V=100$ attempts to modify the volumes.
A  translational/rotational move is defined as a displacement  in each direction of a random quantity distributed uniformly between $\pm ~0.05~\sigma$ and a rotation around a random axis of random angle distributed uniformly  between $\pm 0.1$ radiant. The choice of such a large ratio between 
translation/rotation and swap attempts, $N_{\Delta}/N_N=100$, is dictated by the 
necessity of ensuring a proper equilibration. In the case of particles with short-range and highly directional interactions this choice is relevant, since the probability of inserting a particle with  the correct orientation and position for bonding is significantly reduced as  compared to the case of spherical interactions. 

We also study the model in which the sticky spots are non-regularly distributed on the surface (see Sec.~\ref{sec:model}). In particular we focus on the location of the
critical point, since the values of critical temperature and density for the corresponding ordered arrangement have already been studied~\cite{bian}.
To assess the effect of the randomness on the location of the critical point
we perform  standard Grand Canonical (GCMC) simulations.  In this ensemble, the
chemical potential $\mu$, the temperature $T$ and the volume $V$ are fixed. MC moves include
insertion and deletion of particles as well as particle translation and rotations.
Translational and rotational moves are identical to the one described above for GEMC. 
In  each particle insertion move, a particle with a different realization  of the location of the spots is placed in the box.
GCMC simulations are extremely helpful in the study of the behavior of the system close to the critical point, since they allow for a correct exploration of the large range of densities
experienced by systems  in the vicinity of a critical point.
To locate the critical point we perform simulations at fixed $T$, $\mu$ and $V$, and we tune  $T$ and $\mu$ until the simulated system shows ample density fluctuations,  signaling the proximity to the critical point. Once a reasonable guess of the  critical point in the $(T,\mu)$-plane is reached, we start at least 8 independent GCMC simulations to improve the statistics of the fluctuations in the number of particles $N$ in the box and of the potential energy $E$. 
The location of the critical point is performed through a 
fitting procedure associated to  histogram reweighting~\cite{histrew} and a comparison 
of the fluctuation distribution of the ordering operator 
$\mathcal{M}$ at the critical point with the universal distribution 
characterizing the Ising class~\cite{Wilding_96}. The ordering operator 
$\mathcal{M}$ of the gas-liquid transition is a linear combination 
$\mathcal{M}\sim \rho +s u$, where $\rho$ is the number density, 
$u$ is the energy density of the system, and $s$ is the field mixing parameter. 
Exactly at the critical point, fluctuations of $\mathcal{M}$ are found to 
follow the Ising model universal distribution~\cite{Wilding_96}. 

\section{Numerical Results and Comparison with Wertheim predictions}
We first focus on the effect of patchiness on the phase coexistence region when the particles functionality $f$ is small.
Fig.~\ref{fig:pdW} shows the numerical phase coexistence curves for systems with a number $f=3$,$4$,$5$ of attractive sites geometrically distributed on the particles surfaces. The figure clearly shows a strong reduction of the phase separation region, i.e. an extention of the region of stabilty of the liquid phase  on decreasing $f$.  Similar behavior is shown by the Wertheim predictions, despite the agreement gets worse on increasing $f$.  Hence, both theory and simulations  confirm~\cite{bian,Zacca1} that a region of densities which is not affected by phase separation is  a characteristic of  patchy interacting particles systems. The reduction of the valence is thus  crucial for suppressing the low temperature  ubiquitous process of separation in a dense and dilute solution of particles always observed with spherical potentials. 

Fig.~\ref{fig:pdW} also suggets that the small functionality of the particles makes it possible to observe chains and clusters in long-lived thermodynamic equilibrium. In other words, patchiness offers a way of sampling equilibrium homogeneous states in a large region of intermediate and small densities, where packing is not any longer the leading driving force controlling the structure of the system.  The shrinking  of the unstable region explains why particles interacting via a limited number of functional groups tend to form, at low temperature, open homogeneous structures, which are stabilized by an extended network of long-lived bonds.

To assess if the shape of the coexistence does depend on $f$ we show in 
Fig.~\ref{fig:pdscaled}  the same data of Fig.~\ref{fig:pdW} represented as a function of  reduced variables, $T/T_c$ and $\rho/\rho_c$. 
While the Wertheim theory predicts a  scaled width of the gas-liquid coexistence that shrinks with $f$, numerical data show that, far from the critical point, the curve for $f=3$ appears to be significantly wider than the one for $f=4$ and $5$. Instead, close to the critical point the shape in reduced units
appears to be rather insensitive on $f$ (in agreement with previous findings~\cite{KumarJCP07}). 

Next we focus on the differences between a geometric and a random distribution of the patches and in particular on the $f$ dependence of  $\rho_c$ and $T_c$ in the two different cases. In the disordered case we vary $f$ from 4 to 6. The results of the GCMC simulations are reported in Tab.~\ref{table:total}, together with corresponding quantities previously calculated for the geometric case~\cite{bian}. The results for the critical point location in the two models are also graphically illustrated in  Fig.~\ref{fig:tcphic}. The same trend with $f$ is shown by both models.

It is interesting to observe that, keeping $f$ constant,   $T_c$ and $\rho_c$ both decrease on moving from the geometric to the random arrangement of the sticky sites.   
This decrease suggests that the propagation of the connectivity is less efficient in the 
random patches case, speaking for (i) a waste of bond formation possibilities and/or  (ii)
a failure in the development of long range paths of bonds.  Concerning point (i), we note that a random distribution of patches on the particle surface may introduce correlation in the formation of adjacent bonds. Indeed
the presence of a bonded interaction may induce a screening effect (and hence a decrease in the probability of forming bonds) on  sites closeby located, 
due to  excluded volume interactions. Concerning (ii), we note that a random distribution of sites may also favor the formation of closed loops of bonds, due to a increase in the number of angular possibilities which  satisfy short ring structures,  which are known to suppress the critical phenomena~\cite{Kindt}.
These observation can also explain why the Wertheim theory predictions (which are based on the assumption of both independent bonds and absence of ring structures) are closer to the
geometric case model (see Tab.~\ref{table:teo}).

We note that Tab.~\ref{table:total} and \ref{table:teo} also report the reduced values of the second virial coefficient at the critical point  $B_2^c/B_2^{HS}$. The analytical expression of  $B_2/B_2^{HS}$ is the following
\begin{equation}
\frac{B_2}{B_2^{HS}}= 1 - f^2 \frac{3}{4\pi}  (e^{\beta u_0}-1)\frac{V_b}{\sigma^3} 
\end{equation}
where $B_2^{HS}=2/3 \pi \sigma^3$ is the hard-sphere  virial coefficient.

We also evaluate the bond probability at the critical point, $p_b^c$, on varing $f$  in both the geometric and random patches models and we compare (see Tab.~\ref{table:pb}) the two sets of values with the Wertheim theoretical predictions. As previously observed, the Wertheim predictions are closer to the regular case, even if the theory is insensitive to the patches distribution on the particles surface. We also note that the critical bond probabilities in the geometric model are comparable with the ones recently calculated 
in  Ref.~\cite{FoffiKern} for particles with the same bonding geometry  interacting via the Kern-Frenkel potential~\cite{Kern_03}.  On the other hand, $p_b^c$ for  the random model  is  significantly larger than for the ordered case, supporting   our scenario of a less efficient propagation of connectivity in the random case as compared to the geometric one.


Finally,  we report in Fig.~\ref{fig:rw5} the critical fluctuations distributions  $P(\mathcal{M})$ of the order parameter $\mathcal{M}$ in both the geometric and random patches models with $f=5$. The calculated distributions are compared to the expected fluctuations at the critical point for systems in the Ising universality class~\cite{Wilding_96}. The comparison provides evidence that the transition belongs to the Ising universality class in both studied cases. This is true for each studied value of $f$. The inset of  Fig.~\ref{fig:rw5} shows the corresponding density fluctuations distributions $P(\rho)$ at the estimated $T_c$ and critical chemical potential $\mu_c$. The distribution becomes more asymmetric on decreasing $\rho_c$, signaling an increasing role of the mixing field $s$ (see also Tab.~\ref{table:total}). This means that, at equal $f$, the density fluctuation distributions are more asymmetric in the random case rather than in the geometric one.  


\section{Discussion and Conclusions}

We study the $f$-dependence of the critical behavior in two different patchy models of $f$-functional particles. In both models, the patchy particles are hard-spheres decorated on their surface by a small number of identical  sticky  sites, interacting via a short-range square-well attraction. The difference between the two studied models is the arrangement of the attractive sites on the particles surface. In the first case sites are arranged on a regular structure (see Fig.~1 of Ref.~\cite{bian}) in the same geometry of recently synthesized patchy colloidal particles~\cite{Manoh_03,mohovald,Blaad06}. In the second case, the distribution of the sites is random and different for each particle. 

We compare numerical results and predictions of the thermodynamic perturbation theory developed by Wertheim~\cite{Werth1,Werth2}. This theory assumes  the condition of single-bond per patch and neglects the possibility of forming loops of bonded particles. As previously suggested in Ref.~\cite{Chap2},  the free-energy expression provided by the Wertheim theory can be interpreted
as the free energy of non-interacting clusters.  We show in the Appendix  A
that the corresponding cluster size distribution is the one provided by
Flory and Stockmayer in their seminal work on chemical gelation~\cite{flory,colby}. In this respect,
our study provides an effective expression for  describing the density and temperature dependence of the free energy in  self-assembly of
branched structures and networks. The theory of equilibrium association for systems that form branched structures is receiving  particular attention in the last years~\cite{safran,karlcn,douglas,Workum_06,23,Kindt}, since these systems are found in many technological 
and biomedical applications, as well as in many biological processes.
It is thus  crucial to provide a  general approach for describing the thermodynamics of the branched polymer self-assembly  over the whole range of temperatures, extending to branched system the work developed  in the last decades for the case of self-assembling chains and wormlike micelles~\cite{cates1,Milchev98,milchevdyn}. 

We explicitly solve the Wertheim theory for the chosen site-site interaction and we evaluate lines of specific heat maxima (a signature of the presence of
a specific bonding process) and the   the gas-liquid coexistence lines for $3 \leq f \leq5$.
Thanks to the mapping between the Wertheim theory and the Flory-Stockmayer
approach we also provide expressions for the dependence on $f$ of the percolation line. 
We find that, for all studied $f$, the percolation line merges into the phase separation curve on the left hand side of the critical point, while the intersection between the $C_V^{max}$ line and the coexistence curve moves from the left to the right of the critical point on increasing $f$. Even if  the Wertheim basic assumptions can 
not  be extended to high valence cases, we speculate that, on further increasing $f$, the intersection between coexistence and  $C_V^{max}$ line  will further shift to larger densities.  In this respect the absence of a  $C_V$ maximum in the spherical case
could be due to the fact that the entire  $C_V^{max}$ line lies  in a region of large densities,  made physically inaccessible by  the progressive slowing down of the dynamics on approaching the  glass transition.  Indeed, on increasing $f$, the
patchy potential tends to a spherical  square-well model with analogous range of interaction. For spherical potentials,  it has been shown that the glass line --- which provides the large-density limit of stability of the liquid state~\cite{nuovotartaglia,sastryprl} ---
intersects the coexistence line at a finite temperature and density.
  
The Wertheim  predictions for the  the gas-liquid coexistence curve are compared to
   results of Gibbs Monte Carlo simulations of the regular patches arrangement model. We find that the reduction of the number of patches is accompanied by an enlargement of the  region of stability of the liquid phase in the $(T,\rho)$ plane, confirming the scenario suggested in Ref.~\cite{bian}. 

 Both the Wertheim theory and the
   simulations show that in models of reduced valence, states with $u_0\gg k_{B}T$ can be approached  in equilibrium and reversibly.    Thus in the presence of patchy interactions it becomes possible in a wide range of densities to cool down the system progressively via a sequence of equilibrium homogeneous states.  This is  not possible in  spherically interacting particles for which phase separation always destabilizes the formation of a homogeneous arrested system at low $T$.
   Exploring homogeneous states at low temperature opens the way for
sampling thermodynamic states characterized by bonds with very long lifetime.  
   When the bond-lifetime becomes comparable to the experimental observation time, a dynamic arrest phenomenon at small packing fraction takes place. The reduction of the valence thus makes it feasible to approach dynamic arrest continuously from equilibrium and to generate a state of matter as close as possible to an ideal gel~\cite{genova,zaccajpcm}.   The relation between arrest in limited valence patchy colloidal particles and arrest in strong network forming molecular and atomic liquids have been recently discussed in Ref.~\cite{simone,cristianosilica,statphys,Zacca2}.

Finally, we also study through Grand Canonical Monte Carlo simulations the location of the critical point for disordered arrangements of sites on the hard-sphere surface. Even in this case, we find that  $T_c$ and $\rho_c$ moves towards lower temperatures and densities on decreasing the number of the patches. This confirms that the maximum number of bonds per particle plays an important role in controlling the stability of the liquid phase.  
The fact that the shift with valence of the critical point towards lower temperature and densities  can be accomplished with either a geometric or random arrangement of patches could be particularly significant  to  experimentalists, since it indicates that ordered arrangements of patches are not absolutely necessary to achieve interesting assembly effects.

We observe that, even if the Wertheim theory is insensitive to the arrangement  of the sticky sites, the theoretical predictions for the critical point  location in the phase diagram are closer to the geometric case model rather than the random one, suggesting  that the propagation of the connectivity is less efficient in the random patches case.
 We recall that the theory  is based on the assumption of (i)  independent bonds and  (ii) absence of closed loops of particles.  Hence, the reduced connectivity of the random model, at equal temperature and density,  could be related to  (i) a reduction of bond formation possibilities, induced by a correlation between nearby sites, and/or to (ii) an   increase in possibilities of ring strucures formation, which disfavor 
 the development of branched bonding patterns.

As a side remark, we add in Appendix B the demonstration that, within the Wertheim theory,  when particles interact only via bonds and no hard-core repulsion is present, the thermodynamic stability line (spinodal) coincides with the percolation line.

\section*{Acknowledgements}
We thank Jack F. Douglas and Julio Largo for fruitful and continuous discussions. 
We acknowledge support from  MIUR-Prin and MCRTN-CT-2003-504712.

\section{Appendix A}

Here we provide a physical insight of the Wertheim theory, by discussing 
two equivalent alternative derivations of the Wertheim bond free-energy.
Both derivations assume the system  of associating particles to be formally equivalent to a system of non-interacting clusters, in thermodynamic equilibrium. For simplicity, we assume a De Broglie length $\Lambda=\sigma=1$ in both derivations.

In the first derivation, we assume that the  cluster-size distribution is provided by the Flory-Stockmayer expressions (see Eq.~\ref{eq:flory}), i.e. that 
the system of $N$ $f$-functional associating particles aggregates in clusters 
characterized by the absence of  closed bonding loops. Bonds are also assumed to be 
uncorrelated so that to each bond is associated the same single-bond free energy ${\cal F}_b$.
In the absence of loops, the number of bonds in a cluster of size $n$ is exactly  $(n-1)$, since each new bond adds one new particle. Hence the bond free energy of the cluster is $(n-1) {\cal F}_b$. If clusters do not interact, the  system free energy $F$ can be written as  sum of  the ideal-gas free energy of each distinct bonding topology cluster-type
(accounting for the translational center of mass degrees of freedom) and a sum over the cluster bond free energies. 
Defining $\rho_n^k$ as  the number density of clusters with size $n$ and with bonding topology $k$ one can write
\begin{equation}\label{eqn:Fini}
\frac{\beta F} {V}= \sum_{n}\sum_{k}\rho_n^k \left[\ln \rho_n^k-1\right] + \sum_{n}\sum_{k}\rho_n^k  (n-1) \beta {\cal F}_b.
\end{equation}
Here $V$ is the volume,  $\beta \equiv 1/k_B T$ (with $k_B$ the Boltzmann constant)
and the sum on $n$ runs over all the possible cluster sizes, from one (monomers) to infinity,  while the sum over $k$ includes all $\omega_n$ distinct  bonding topology of clusters of size $n$.  Since clusters with the same size but different bonding pattern are equiprobable, then $\rho_n^k \equiv \rho_n / \omega_n$  (see Eq.~\ref{eq:flory}). Thus Eq.~\ref{eqn:Fini} becomes
\begin{equation}
\label{eq:Fmiddle}
\frac{\beta F} {V}= \sum_n \rho_n \left[\ln \frac{\rho_n}{\omega_n}-1\right] + \sum_{n}\rho_n (n-1) \beta {\cal F}_b.
\end{equation}
Substituting Eq.~\ref{eq:flory} in Eq.~\ref{eq:Fmiddle} and summing over $n$ one obtains, for $p_b < (f-1)^{-1}$ (which express the condition that all clusters are finite~\cite{flory}),  the following expression for the system free energy in term of $p_b$ and bond free energy
\begin{eqnarray}
\frac{\beta F} {V} &=& \rho \ln\rho - \rho + \\
\nonumber
&+& \rho \ln (1-p_b)^f -  \frac{\rho f p_b}{2}  \left[  \ln    \frac{\rho (1-p_b)^2}{p_b} e^{- \beta {\cal F}_b}  -1\right].
\end{eqnarray}
Such expression can be seen as a high temperature contribution~\cite{Hansennew} ($\rho \ln\rho - \rho$) plus a remaining bonding term.
The bonding free energy coincides with the Wertheim expression~\cite{Werth1,Werth2,Hansennew}  ($\rho \ln (1-p_b)^f + \frac{\rho f p_b}{2} $)  when  the connection between 
 $p_b$ and $  {\cal F}_b$ is given by Eq.~\ref{eq:pb}.
 This simple derivation can be also extended to binary mixtures.

An even simpler derivation has been suggested in Ref.~\cite{Chap2} and it is here reported for
completeness. Also this derivation assumes that the system is an ideal gas of clusters
and hence that the product $\beta PV$ is identical to the number of clusters $N_c$. 
The evaluation of $N_c$ is straightforward for $p_b$ values smaller than the percolation threshold $p_b^p$, since, in the absence of closed bond loops, $N_{c}$ is equal to the  number of particles  minus the number of bonds $N_b$.
Calling $N_b^{max}= \frac{Nf}{2}$ the maximum number of possible bonds and noting that
 $p_b$ is the ratio between $N_b$ and  $N_b^{max}$,  one finds 
\begin{equation}
N_c=N-N_b=N \left(1 - \frac{f}{2} p_b\right )
\label{eq:nc}
\end{equation}
and
\begin{equation}
\beta P  = \rho \left(1 - \frac{f}{2} p_b\right ).
\label{eq:pvnc}
\end{equation}
Since  the system is in dynamic equilibrium, the particle chemical potential is independent from the
cluster to which the particle belongs to. Hence, it is identical to the chemical potential $\mu$ of the monomer. The ideal-gas hypothesis  implies that the activity of the monomer  $z\equiv \exp(\beta \mu)$ is related to the monomer number density by $z=\rho_1=\rho (1-p_b)^f$.  Hence
the system free energy density  can be immediately written as
\begin{equation}
\beta F = \rho \beta \mu -\beta P    = \rho \ln \left[\rho (1-p_b)^f\right ] - \rho \left(1 - \frac{f}{2} p_b\right )
\end{equation}
which coincides with the Wertheim expression when the reference system is the ideal gas.

We note in conclusions that the above relations are valid only before percolation. Indeed,
the sums over $n$ in Eq.~\ref{eq:Fmiddle} as well as  Eq.~\ref{eq:nc} are valid only for $p_b< p_b^p$.
Hence, the region of validity of  the Wertheim theory should be strictly limited to non-percolating states. Nevertheless we observe that the theory   works well even below the percolation threshold~\cite{bian,23}.   It could be in principle possible to extend the formalism to $p_b> p_b^p$ by accounting correctly for the $p_b$ dependence of the number of clusters (which is always possible, analytically for small $f$ and numerically above), but it is not clear how to handle the free energy contribution of the percolating cluster.

\section{Appendix B}

In this Appendix we examine the thermodynamic stability of a system composed of non-interacting
clusters, described by the free energy of Eq.~\ref{eq:Fmiddle}.  A stable system is characterized
by a negative 
volume derivative of the pressure. The region of stability is delimited by the so-called spinodal line, defined as the locus of points such that $\left(\partial \beta P/\partial V\right)_T =0$.
The volume derivative of the pressure, under the ideal gas approximation $g_{HS}(r) = 1$, is controlled only by the volume derivative of $p_b$ (see Eq.~\ref{eq:pvnc}). Interestingly, it gives for the bond probability at the spinodal line $p_b^s$ 
\begin{equation}
p_b^s=\frac{1}{f-1}
\end{equation}
i.e. the same condition that defines percolation. Hence $p_b^s=p_b^p$.  The system is thus  mechanically stable  only in the non-percolating region.  In other words in the ideal gas approximation no dense stable states are possible and the system exists only in the gas-phase.  In the Wertheim theory the existence of a liquid phase is generated by the significant increase of the pressure at low $V$ introduced by the hard-sphere reference contribution. Fig.~\ref{fig:p} provides an example of the effect of the hard-sphere contribution on the pressure for the case $f=3$.

We conclude noting that the absence of the hard-core repulsion appears to be essential in formally associating the percolation line with the spinodal line, providing an analytic simple example of the possibility of interpreting critical phenomena in term of percolation~\cite{coniglio,padua}.

\bibliographystyle{apsrev}
\bibliography{./A712051JCP.bib}

\begin{thebibliography}{70}
\expandafter\ifx\csname natexlab\endcsname\relax\def\natexlab#1{#1}\fi
\expandafter\ifx\csname bibnamefont\endcsname\relax
  \def\bibnamefont#1{#1}\fi
\expandafter\ifx\csname bibfnamefont\endcsname\relax
  \def\bibfnamefont#1{#1}\fi
\expandafter\ifx\csname citenamefont\endcsname\relax
  \def\citenamefont#1{#1}\fi
\expandafter\ifx\csname url\endcsname\relax
  \def\url#1{\texttt{#1}}\fi
\expandafter\ifx\csname urlprefix\endcsname\relax\def\urlprefix{URL }\fi
\providecommand{\bibinfo}[2]{#2}
\providecommand{\eprint}[2][]{\url{#2}}

\bibitem[{\citenamefont{Trappe and Sandk\"uhler}(2004)}]{Trappe}
\bibinfo{author}{\bibfnamefont{V.}~\bibnamefont{Trappe}} \bibnamefont{and}
  \bibinfo{author}{\bibfnamefont{P.}~\bibnamefont{Sandk\"uhler}},
  \bibinfo{journal}{Curr. Op. Colloid. Interf. Sci.}
  \textbf{\bibinfo{volume}{8}}, \bibinfo{pages}{494} (\bibinfo{year}{2004}).

\bibitem[{\citenamefont{Cipelletti and Ramos}(2005)}]{Cipelletti}
\bibinfo{author}{\bibfnamefont{L.}~\bibnamefont{Cipelletti}} \bibnamefont{and}
  \bibinfo{author}{\bibfnamefont{L.}~\bibnamefont{Ramos}}, \bibinfo{journal}{J.
  Phys.: Condens. Matter} \textbf{\bibinfo{volume}{17}}, \bibinfo{pages}{253}
  (\bibinfo{year}{2005}).

\bibitem[{\citenamefont{Zaccarelli}(2007)}]{zaccajpcm}
\bibinfo{author}{\bibfnamefont{E.}~\bibnamefont{Zaccarelli}},
  \bibinfo{journal}{J. Phys.: Condens. Matter} \textbf{\bibinfo{volume}{19}},
  \bibinfo{pages}{323101} (\bibinfo{year}{2007}).

\bibitem[{\citenamefont{Sciortino and Tartaglia}(2005)}]{advances}
\bibinfo{author}{\bibfnamefont{F.}~\bibnamefont{Sciortino}} \bibnamefont{and}
  \bibinfo{author}{\bibfnamefont{P.}~\bibnamefont{Tartaglia}},
  \bibinfo{journal}{Adv. Phys.} \textbf{\bibinfo{volume}{54}},
  \bibinfo{pages}{471} (\bibinfo{year}{2005}).

\bibitem[{\citenamefont{Sciortino et~al.}(2005)\citenamefont{Sciortino,
  Buldyrev, {De~Michele}, Ghofraniha, {La Nave}, Moreno, Mossa, Tartaglia, and
  Zaccarelli}}]{genova}
\bibinfo{author}{\bibfnamefont{F.}~\bibnamefont{Sciortino}},
  \bibinfo{author}{\bibfnamefont{S.}~\bibnamefont{Buldyrev}},
  \bibinfo{author}{\bibfnamefont{C.}~\bibnamefont{{De~Michele}}},
  \bibinfo{author}{\bibfnamefont{N.}~\bibnamefont{Ghofraniha}},
  \bibinfo{author}{\bibfnamefont{E.}~\bibnamefont{{La Nave}}},
  \bibinfo{author}{\bibfnamefont{A.}~\bibnamefont{Moreno}},
  \bibinfo{author}{\bibfnamefont{S.}~\bibnamefont{Mossa}},
  \bibinfo{author}{\bibfnamefont{P.}~\bibnamefont{Tartaglia}},
  \bibnamefont{and}
  \bibinfo{author}{\bibfnamefont{E.}~\bibnamefont{Zaccarelli}},
  \bibinfo{journal}{Comp. Phys. Comm.} \textbf{\bibinfo{volume}{169}},
  \bibinfo{pages}{166} (\bibinfo{year}{2005}).

\bibitem[{\citenamefont{Zaccarelli et~al.}(2005)\citenamefont{Zaccarelli,
  Buldyrev, {La Nave}, Moreno, Saika-Voivod, Sciortino, and
  Tartaglia}}]{Zacca1}
\bibinfo{author}{\bibfnamefont{E.}~\bibnamefont{Zaccarelli}},
  \bibinfo{author}{\bibfnamefont{S.~V.} \bibnamefont{Buldyrev}},
  \bibinfo{author}{\bibfnamefont{E.}~\bibnamefont{{La Nave}}},
  \bibinfo{author}{\bibfnamefont{A.~J.} \bibnamefont{Moreno}},
  \bibinfo{author}{\bibnamefont{Saika-Voivod}},
  \bibinfo{author}{\bibfnamefont{F.}~\bibnamefont{Sciortino}},
  \bibnamefont{and}
  \bibinfo{author}{\bibfnamefont{P.}~\bibnamefont{Tartaglia}},
  \bibinfo{journal}{Phys. Rev. Lett.} \textbf{\bibinfo{volume}{{\bf 94}}},
  \bibinfo{pages}{218301} (\bibinfo{year}{2005}).

\bibitem[{\citenamefont{Bianchi et~al.}(2006)\citenamefont{Bianchi, Largo,
  Tartaglia, Zaccarelli, and Sciortino}}]{bian}
\bibinfo{author}{\bibfnamefont{E.}~\bibnamefont{Bianchi}},
  \bibinfo{author}{\bibfnamefont{J.}~\bibnamefont{Largo}},
  \bibinfo{author}{\bibfnamefont{P.}~\bibnamefont{Tartaglia}},
  \bibinfo{author}{\bibfnamefont{E.}~\bibnamefont{Zaccarelli}},
  \bibnamefont{and}
  \bibinfo{author}{\bibfnamefont{F.}~\bibnamefont{Sciortino}},
  \bibinfo{journal}{Phys. Rev. Lett.} \textbf{\bibinfo{volume}{97}},
  \bibinfo{pages}{168301} (\bibinfo{year}{2006}).

\bibitem[{\citenamefont{Wilber et~al.}(2007)\citenamefont{Wilber, Doye, Louis,
  Noya, Miller, and Wong}}]{Wilber-2006}
\bibinfo{author}{\bibfnamefont{A.~W.} \bibnamefont{Wilber}},
  \bibinfo{author}{\bibfnamefont{J.~P.~K.} \bibnamefont{Doye}},
  \bibinfo{author}{\bibfnamefont{A.~A.} \bibnamefont{Louis}},
  \bibinfo{author}{\bibfnamefont{E.~G.} \bibnamefont{Noya}},
  \bibinfo{author}{\bibfnamefont{M.~A.} \bibnamefont{Miller}},
  \bibnamefont{and} \bibinfo{author}{\bibfnamefont{P.}~\bibnamefont{Wong}},
  \bibinfo{journal}{J. Chem. Phys.} \textbf{\bibinfo{volume}{127}},
  \bibinfo{pages}{085106} (\bibinfo{year}{2007}).

\bibitem[{\citenamefont{Doye et~al.}(2007)\citenamefont{Doye, Louis, Lin,
  Allen, Noya, Wilber, Kok, and Lyus}}]{doye}
\bibinfo{author}{\bibfnamefont{J.~P.~K.} \bibnamefont{Doye}},
  \bibinfo{author}{\bibfnamefont{A.~A.} \bibnamefont{Louis}},
  \bibinfo{author}{\bibfnamefont{I.-C.} \bibnamefont{Lin}},
  \bibinfo{author}{\bibfnamefont{L.~R.} \bibnamefont{Allen}},
  \bibinfo{author}{\bibfnamefont{E.~G.} \bibnamefont{Noya}},
  \bibinfo{author}{\bibfnamefont{A.~W.} \bibnamefont{Wilber}},
  \bibinfo{author}{\bibfnamefont{H.~C.} \bibnamefont{Kok}}, \bibnamefont{and}
  \bibinfo{author}{\bibfnamefont{R.}~\bibnamefont{Lyus}},
  \bibinfo{journal}{Phys. Chem. Chem. Phys.} \textbf{\bibinfo{volume}{9}},
  \bibinfo{pages}{2197} (\bibinfo{year}{2007}).

\bibitem[{\citenamefont{Noya et~al.}(2007)\citenamefont{Noya, Vega, Doye, and
  Louis}}]{octahedral}
\bibinfo{author}{\bibfnamefont{E.~G.} \bibnamefont{Noya}},
  \bibinfo{author}{\bibfnamefont{C.}~\bibnamefont{Vega}},
  \bibinfo{author}{\bibfnamefont{J.~P.~K.} \bibnamefont{Doye}},
  \bibnamefont{and} \bibinfo{author}{\bibfnamefont{A.~A.} \bibnamefont{Louis}},
  \bibinfo{journal}{J. Chem. Phys.} \textbf{\bibinfo{volume}{127}},
  \bibinfo{pages}{054501} (\bibinfo{year}{2007}).

\bibitem[{\citenamefont{Glotzer and Solomon}(2007)}]{Glotz_Solomon_natmat}
\bibinfo{author}{\bibfnamefont{S.~C.} \bibnamefont{Glotzer}} \bibnamefont{and}
  \bibinfo{author}{\bibfnamefont{M.~J.} \bibnamefont{Solomon}},
  \bibinfo{journal}{Nat. Mat.} \textbf{\bibinfo{volume}{6}},
  \bibinfo{pages}{557} (\bibinfo{year}{2007}).

\bibitem[{\citenamefont{Zhang and Glotzer}(2004)}]{zhangglotzer}
\bibinfo{author}{\bibfnamefont{Z.}~\bibnamefont{Zhang}} \bibnamefont{and}
  \bibinfo{author}{\bibfnamefont{S.~C.} \bibnamefont{Glotzer}},
  \bibinfo{journal}{Nanoletters} \textbf{\bibinfo{volume}{4}},
  \bibinfo{pages}{1407} (\bibinfo{year}{2004}).

\bibitem[{\citenamefont{Zhang et~al.}(2003)\citenamefont{Zhang, Horsch, Lamm,
  and Glotzer}}]{Zhang_03}
\bibinfo{author}{\bibfnamefont{Z.}~\bibnamefont{Zhang}},
  \bibinfo{author}{\bibfnamefont{M.~A.} \bibnamefont{Horsch}},
  \bibinfo{author}{\bibfnamefont{M.~H.} \bibnamefont{Lamm}}, \bibnamefont{and}
  \bibinfo{author}{\bibfnamefont{S.~C.} \bibnamefont{Glotzer}},
  \bibinfo{journal}{Nano Lett.} \textbf{\bibinfo{volume}{{\bf 3}}},
  \bibinfo{pages}{1341} (\bibinfo{year}{2003}).

\bibitem[{\citenamefont{{De~Michele}
  et~al.}(2006{\natexlab{a}})\citenamefont{{De~Michele}, Gabrielli, Tartaglia,
  and Sciortino}}]{simone}
\bibinfo{author}{\bibfnamefont{C.}~\bibnamefont{{De~Michele}}},
  \bibinfo{author}{\bibfnamefont{S.}~\bibnamefont{Gabrielli}},
  \bibinfo{author}{\bibfnamefont{P.}~\bibnamefont{Tartaglia}},
  \bibnamefont{and}
  \bibinfo{author}{\bibfnamefont{F.}~\bibnamefont{Sciortino}},
  \bibinfo{journal}{J. Phys. Chem. B} \textbf{\bibinfo{volume}{110}},
  \bibinfo{pages}{8064 } (\bibinfo{year}{2006}{\natexlab{a}}).

\bibitem[{\citenamefont{Huisman et~al.}(2007)\citenamefont{Huisman, Bolhuis,
  and Fasolino}}]{bastiaan}
\bibinfo{author}{\bibfnamefont{B.~A.~H.} \bibnamefont{Huisman}},
  \bibinfo{author}{\bibfnamefont{P.~G.} \bibnamefont{Bolhuis}},
  \bibnamefont{and} \bibinfo{author}{\bibfnamefont{A.}~\bibnamefont{Fasolino}}
  (\bibinfo{year}{2007}), \eprint{cond-mat/0711.4704}.

\bibitem[{\citenamefont{Manoharan et~al.}(2003)\citenamefont{Manoharan,
  Elsesser, and Pine}}]{Manoh_03}
\bibinfo{author}{\bibfnamefont{V.~N.} \bibnamefont{Manoharan}},
  \bibinfo{author}{\bibfnamefont{M.~T.} \bibnamefont{Elsesser}},
  \bibnamefont{and} \bibinfo{author}{\bibfnamefont{D.~J.} \bibnamefont{Pine}},
  \bibinfo{journal}{Science} \textbf{\bibinfo{volume}{{301}}},
  \bibinfo{pages}{483} (\bibinfo{year}{2003}).

\bibitem[{\citenamefont{Zhang et~al.}(2005)\citenamefont{Zhang, Wang, and
  M\"ohwald}}]{mohovald}
\bibinfo{author}{\bibfnamefont{G.}~\bibnamefont{Zhang}},
  \bibinfo{author}{\bibfnamefont{D.}~\bibnamefont{Wang}}, \bibnamefont{and}
  \bibinfo{author}{\bibfnamefont{H.}~\bibnamefont{M\"ohwald}},
  \bibinfo{journal}{Angew. Chem. Int. Ed.} \textbf{\bibinfo{volume}{44}},
  \bibinfo{pages}{1} (\bibinfo{year}{2005}).

\bibitem[{\citenamefont{van Blaaderen}(2006)}]{Blaad06}
\bibinfo{author}{\bibfnamefont{A.}~\bibnamefont{van Blaaderen}},
  \bibinfo{journal}{News and Views, Nature} \textbf{\bibinfo{volume}{439}},
  \bibinfo{pages}{545} (\bibinfo{year}{2006}).

\bibitem[{\citenamefont{Cho et~al.}(2005)\citenamefont{Cho, Yi, Lim, Kim,
  Manoharan, Pine, and Yang}}]{Cho_05}
\bibinfo{author}{\bibfnamefont{Y.-S.} \bibnamefont{Cho}},
  \bibinfo{author}{\bibfnamefont{G.-R.} \bibnamefont{Yi}},
  \bibinfo{author}{\bibfnamefont{J.-M.} \bibnamefont{Lim}},
  \bibinfo{author}{\bibfnamefont{S.-H.} \bibnamefont{Kim}},
  \bibinfo{author}{\bibfnamefont{V.~N.} \bibnamefont{Manoharan}},
  \bibinfo{author}{\bibfnamefont{D.~J.} \bibnamefont{Pine}}, \bibnamefont{and}
  \bibinfo{author}{\bibfnamefont{S.-M.} \bibnamefont{Yang}},
  \bibinfo{journal}{J. Am. Chem. Soc.} \textbf{\bibinfo{volume}{127}},
  \bibinfo{pages}{15968} (\bibinfo{year}{2005}).

\bibitem[{\citenamefont{Glotzer et~al.}(2004)\citenamefont{Glotzer, Solomon,
  and Kotov}}]{Glotz_Solomon}
\bibinfo{author}{\bibfnamefont{S.~C.} \bibnamefont{Glotzer}},
  \bibinfo{author}{\bibfnamefont{M.~J.} \bibnamefont{Solomon}},
  \bibnamefont{and} \bibinfo{author}{\bibfnamefont{N.~A.} \bibnamefont{Kotov}},
  \bibinfo{journal}{AIChE Journal} \textbf{\bibinfo{volume}{{\bf 50}}},
  \bibinfo{pages}{2978} (\bibinfo{year}{2004}).

\bibitem[{\citenamefont{Leibler}(2005)}]{leibler}
\bibinfo{author}{\bibfnamefont{L.}~\bibnamefont{Leibler}},
  \bibinfo{journal}{Progress in Polymer Science} \textbf{\bibinfo{volume}{30}},
  \bibinfo{pages}{898} (\bibinfo{year}{2005}).

\bibitem[{\citenamefont{{Lehn}}(2002{\natexlab{a}})}]{lehn1}
\bibinfo{author}{\bibfnamefont{J.-M.} \bibnamefont{{Lehn}}},
  \bibinfo{journal}{Science} \textbf{\bibinfo{volume}{295}},
  \bibinfo{pages}{2400} (\bibinfo{year}{2002}{\natexlab{a}}).

\bibitem[{\citenamefont{{Lehn}}(2002{\natexlab{b}})}]{lehn2}
\bibinfo{author}{\bibfnamefont{J.-M.} \bibnamefont{{Lehn}}},
  \bibinfo{journal}{Proc. Natl. Acad. Sci.} \textbf{\bibinfo{volume}{99}},
  \bibinfo{pages}{4763} (\bibinfo{year}{2002}{\natexlab{b}}).

\bibitem[{\citenamefont{Lomakin et~al.}(1999)\citenamefont{Lomakin, Asherie,
  and Benedek}}]{Lomakin}
\bibinfo{author}{\bibfnamefont{A.}~\bibnamefont{Lomakin}},
  \bibinfo{author}{\bibfnamefont{N.}~\bibnamefont{Asherie}}, \bibnamefont{and}
  \bibinfo{author}{\bibfnamefont{G.}~\bibnamefont{Benedek}},
  \bibinfo{journal}{Proc. Natl. Acad. Sci.} \textbf{\bibinfo{volume}{96}},
  \bibinfo{pages}{9465} (\bibinfo{year}{1999}).

\bibitem[{\citenamefont{Sear}(1999)}]{Sear_99}
\bibinfo{author}{\bibfnamefont{R.~P.} \bibnamefont{Sear}}, \bibinfo{journal}{J.
  Chem. Phys.} \textbf{\bibinfo{volume}{111}}, \bibinfo{pages}{4800}
  (\bibinfo{year}{1999}).

\bibitem[{\citenamefont{Liu et~al.}(2007)\citenamefont{Liu, Kumar, and
  Sciortino}}]{KumarJCP07}
\bibinfo{author}{\bibfnamefont{H.}~\bibnamefont{Liu}},
  \bibinfo{author}{\bibfnamefont{S.~K.} \bibnamefont{Kumar}}, \bibnamefont{and}
  \bibinfo{author}{\bibfnamefont{F.}~\bibnamefont{Sciortino}},
  \bibinfo{journal}{J. Chem. Phys.} \textbf{\bibinfo{volume}{127}},
  \bibinfo{pages}{084902} (\bibinfo{year}{2007}).

\bibitem[{\citenamefont{McManus et~al.}(2007)\citenamefont{McManus, Lomakin,
  Basan, Pande, Ogun, Pande, and Benedek}}]{McManus}
\bibinfo{author}{\bibfnamefont{J.~J.} \bibnamefont{McManus}},
  \bibinfo{author}{\bibfnamefont{A.}~\bibnamefont{Lomakin}},
  \bibinfo{author}{\bibfnamefont{M.}~\bibnamefont{Basan}},
  \bibinfo{author}{\bibfnamefont{A.}~\bibnamefont{Pande}},
  \bibinfo{author}{\bibfnamefont{O.}~\bibnamefont{Ogun}},
  \bibinfo{author}{\bibfnamefont{J.}~\bibnamefont{Pande}}, \bibnamefont{and}
  \bibinfo{author}{\bibfnamefont{G.~B.} \bibnamefont{Benedek}},
  \bibinfo{journal}{Proc. Natl. Acad. Sci.} \textbf{\bibinfo{volume}{104}},
  \bibinfo{pages}{16856} (\bibinfo{year}{2007}).

\bibitem[{\citenamefont{Wertheim}(1984)}]{Werth1}
\bibinfo{author}{\bibfnamefont{M.}~\bibnamefont{Wertheim}},
  \bibinfo{journal}{J. Stat. Phys.} \textbf{\bibinfo{volume}{35}},
  \bibinfo{pages}{19, ibid. 35} (\bibinfo{year}{1984}).

\bibitem[{\citenamefont{Wertheim}(1986{\natexlab{a}})}]{Werth2}
\bibinfo{author}{\bibfnamefont{M.}~\bibnamefont{Wertheim}},
  \bibinfo{journal}{J. Stat. Phys.} \textbf{\bibinfo{volume}{42}},
  \bibinfo{pages}{459, ibid. 477} (\bibinfo{year}{1986}{\natexlab{a}}).

\bibitem[{\citenamefont{Mirkin et~al.}(1996)\citenamefont{Mirkin, Letsinger,
  Mucic, and Storhoff.}}]{dna}
\bibinfo{author}{\bibfnamefont{C.}~\bibnamefont{Mirkin}},
  \bibinfo{author}{\bibfnamefont{R.}~\bibnamefont{Letsinger}},
  \bibinfo{author}{\bibfnamefont{R.}~\bibnamefont{Mucic}}, \bibnamefont{and}
  \bibinfo{author}{\bibfnamefont{J.}~\bibnamefont{Storhoff.}},
  \bibinfo{journal}{Nature} \textbf{\bibinfo{volume}{382}},
  \bibinfo{pages}{607} (\bibinfo{year}{1996}).

\bibitem[{\citenamefont{{Starr} and {Sciortino}}(2006)}]{dnastarr}
\bibinfo{author}{\bibfnamefont{F.~W.} \bibnamefont{{Starr}}} \bibnamefont{and}
  \bibinfo{author}{\bibfnamefont{F.}~\bibnamefont{{Sciortino}}},
  \bibinfo{journal}{J. Phys.: Condens. Matter} \textbf{\bibinfo{volume}{18}},
  \bibinfo{pages}{L347} (\bibinfo{year}{2006}).

\bibitem[{\citenamefont{Hiddessen et~al.}(2000)\citenamefont{Hiddessen,
  Rotgers, Weitz, and Hammer}}]{hiddessen1}
\bibinfo{author}{\bibfnamefont{A.~L.} \bibnamefont{Hiddessen}},
  \bibinfo{author}{\bibfnamefont{S.~D.} \bibnamefont{Rotgers}},
  \bibinfo{author}{\bibfnamefont{D.~A.} \bibnamefont{Weitz}}, \bibnamefont{and}
  \bibinfo{author}{\bibfnamefont{D.~A.} \bibnamefont{Hammer}},
  \bibinfo{journal}{Langmuir} \textbf{\bibinfo{volume}{16}},
  \bibinfo{pages}{9744} (\bibinfo{year}{2000}).

\bibitem[{\citenamefont{Hiddessen et~al.}(2004)\citenamefont{Hiddessen, Weitz,
  and Hammer}}]{hiddessen2}
\bibinfo{author}{\bibfnamefont{A.~L.} \bibnamefont{Hiddessen}},
  \bibinfo{author}{\bibfnamefont{D.~A.} \bibnamefont{Weitz}}, \bibnamefont{and}
  \bibinfo{author}{\bibfnamefont{D.~A.} \bibnamefont{Hammer}},
  \bibinfo{journal}{Langmuir} \textbf{\bibinfo{volume}{20}},
  \bibinfo{pages}{71} (\bibinfo{year}{2004}).

\bibitem[{\citenamefont{Jackson et~al.}(1988)\citenamefont{Jackson, Chapman,
  and Gubbins}}]{Chap1}
\bibinfo{author}{\bibfnamefont{G.}~\bibnamefont{Jackson}},
  \bibinfo{author}{\bibfnamefont{W.~G.} \bibnamefont{Chapman}},
  \bibnamefont{and} \bibinfo{author}{\bibfnamefont{K.~E.}
  \bibnamefont{Gubbins}}, \bibinfo{journal}{Mol. Phys.}
  \textbf{\bibinfo{volume}{{\bf 65}}}, \bibinfo{pages}{1}
  (\bibinfo{year}{1988}).

\bibitem[{\citenamefont{Chapman et~al.}(1988)\citenamefont{Chapman, Jackson,
  and Gubbins}}]{Chap2}
\bibinfo{author}{\bibfnamefont{G.}~\bibnamefont{Chapman}},
  \bibinfo{author}{\bibfnamefont{G.}~\bibnamefont{Jackson}}, \bibnamefont{and}
  \bibinfo{author}{\bibfnamefont{K.~E.} \bibnamefont{Gubbins}},
  \bibinfo{journal}{Mol. Phys.} \textbf{\bibinfo{volume}{{\bf 65}}},
  \bibinfo{pages}{1057} (\bibinfo{year}{1988}).

\bibitem[{\citenamefont{Gil-Villegas et~al.}(1997)\citenamefont{Gil-Villegas,
  Galindo, Whitehead, Mills, Jackson, and Burgess}}]{Saft-vr}
\bibinfo{author}{\bibfnamefont{A.}~\bibnamefont{Gil-Villegas}},
  \bibinfo{author}{\bibfnamefont{A.}~\bibnamefont{Galindo}},
  \bibinfo{author}{\bibfnamefont{P.~J.} \bibnamefont{Whitehead}},
  \bibinfo{author}{\bibfnamefont{S.~J.} \bibnamefont{Mills}},
  \bibinfo{author}{\bibfnamefont{G.}~\bibnamefont{Jackson}}, \bibnamefont{and}
  \bibinfo{author}{\bibfnamefont{A.~N.} \bibnamefont{Burgess}},
  \bibinfo{journal}{J. Chem. Phys.} \textbf{\bibinfo{volume}{106}},
  \bibinfo{pages}{4168} (\bibinfo{year}{1997}).

\bibitem[{\citenamefont{Galindo et~al.}(1999)\citenamefont{Galindo,
  Gil-Villegas, Jackson, and Burgess}}]{Saft-vre}
\bibinfo{author}{\bibfnamefont{A.}~\bibnamefont{Galindo}},
  \bibinfo{author}{\bibfnamefont{A.}~\bibnamefont{Gil-Villegas}},
  \bibinfo{author}{\bibfnamefont{G.}~\bibnamefont{Jackson}}, \bibnamefont{and}
  \bibinfo{author}{\bibfnamefont{A.~N.} \bibnamefont{Burgess}},
  \bibinfo{journal}{J. Phys. Chem.} \textbf{\bibinfo{volume}{103}},
  \bibinfo{pages}{10272} (\bibinfo{year}{1999}).

\bibitem[{\citenamefont{{Bianchi} et~al.}(2007)\citenamefont{{Bianchi},
  {Tartaglia}, {La Nave}, and {Sciortino}}}]{23}
\bibinfo{author}{\bibfnamefont{E.}~\bibnamefont{{Bianchi}}},
  \bibinfo{author}{\bibfnamefont{P.}~\bibnamefont{{Tartaglia}}},
  \bibinfo{author}{\bibfnamefont{E.}~\bibnamefont{{La Nave}}},
  \bibnamefont{and}
  \bibinfo{author}{\bibfnamefont{F.}~\bibnamefont{{Sciortino}}},
  \bibinfo{journal}{J. Phys. Chem. B} \textbf{\bibinfo{volume}{111}},
  \bibinfo{pages}{11765} (\bibinfo{year}{2007}).

\bibitem[{\citenamefont{Flory}(1953)}]{flory}
\bibinfo{author}{\bibfnamefont{P.~J.} \bibnamefont{Flory}},
  \emph{\bibinfo{title}{Principles of polymer chemistry}}
  (\bibinfo{publisher}{Cornell University Press (Ithaca and London)},
  \bibinfo{year}{1953}).

\bibitem[{\citenamefont{Hansen and McDonald}(2006)}]{Hansennew}
\bibinfo{author}{\bibfnamefont{J.~P.} \bibnamefont{Hansen}} \bibnamefont{and}
  \bibinfo{author}{\bibfnamefont{I.~R.} \bibnamefont{McDonald}},
  \emph{\bibinfo{title}{Theory of simple liquids}}
  (\bibinfo{publisher}{Academic Press, New York}, \bibinfo{year}{2006}),
  \bibinfo{edition}{3rd} ed.

\bibitem[{\citenamefont{Wertheim}(1986{\natexlab{b}})}]{Werth5}
\bibinfo{author}{\bibfnamefont{M.}~\bibnamefont{Wertheim}},
  \bibinfo{journal}{J. Chem. Phys.} \textbf{\bibinfo{volume}{{ 85}}},
  \bibinfo{pages}{2929} (\bibinfo{year}{1986}{\natexlab{b}}).

\bibitem[{\citenamefont{Nezbeda and Iglesia-Silva}(1990)}]{Nez_90}
\bibinfo{author}{\bibfnamefont{I.}~\bibnamefont{Nezbeda}} \bibnamefont{and}
  \bibinfo{author}{\bibfnamefont{G.}~\bibnamefont{Iglesia-Silva}},
  \bibinfo{journal}{Mol. Phys.} \textbf{\bibinfo{volume}{{\bf 69}}},
  \bibinfo{pages}{767} (\bibinfo{year}{1990}).

\bibitem[{\citenamefont{Carnahan and Starling}(1969)}]{CS_69}
\bibinfo{author}{\bibfnamefont{N.~F.} \bibnamefont{Carnahan}} \bibnamefont{and}
  \bibinfo{author}{\bibfnamefont{K.~E.} \bibnamefont{Starling}},
  \bibinfo{journal}{J. Chem. Phys.} \textbf{\bibinfo{volume}{{\bf 51}}},
  \bibinfo{pages}{635} (\bibinfo{year}{1969}).

\bibitem[{\citenamefont{Sciortino et~al.}(2007)\citenamefont{Sciortino,
  Bianchi, F.Douglas, and Tartaglia}}]{M2}
\bibinfo{author}{\bibfnamefont{F.}~\bibnamefont{Sciortino}},
  \bibinfo{author}{\bibfnamefont{E.}~\bibnamefont{Bianchi}},
  \bibinfo{author}{\bibfnamefont{J.}~\bibnamefont{F.Douglas}},
  \bibnamefont{and}
  \bibinfo{author}{\bibfnamefont{P.}~\bibnamefont{Tartaglia}},
  \bibinfo{journal}{J. Chem. Phys.} \textbf{\bibinfo{volume}{126}},
  \bibinfo{pages}{194903} (\bibinfo{year}{2007}).

\bibitem[{\citenamefont{Greer}(1988)}]{Greer88}
\bibinfo{author}{\bibfnamefont{S.~C.} \bibnamefont{Greer}},
  \bibinfo{journal}{J. Phys. Chem. B} \textbf{\bibinfo{volume}{102}},
  \bibinfo{pages}{5413} (\bibinfo{year}{1988}).

\bibitem[{\citenamefont{Greer}(1996)}]{Greer96}
\bibinfo{author}{\bibfnamefont{S.~C.} \bibnamefont{Greer}},
  \bibinfo{journal}{Adv. Chem. Phys.} \textbf{\bibinfo{volume}{94}},
  \bibinfo{pages}{261} (\bibinfo{year}{1996}).

\bibitem[{\citenamefont{Greer}(2002)}]{Greer02}
\bibinfo{author}{\bibfnamefont{S.~C.} \bibnamefont{Greer}},
  \bibinfo{journal}{Ann. Rev. Phys. Chem.} \textbf{\bibinfo{volume}{53}},
  \bibinfo{pages}{173} (\bibinfo{year}{2002}).

\bibitem[{\citenamefont{{Wittmer} et~al.}(1998)\citenamefont{{Wittmer},
  {Milchev}, and {Cates}}}]{Milchev98}
\bibinfo{author}{\bibfnamefont{J.~P.} \bibnamefont{{Wittmer}}},
  \bibinfo{author}{\bibfnamefont{A.}~\bibnamefont{{Milchev}}},
  \bibnamefont{and} \bibinfo{author}{\bibfnamefont{M.~E.}
  \bibnamefont{{Cates}}}, \bibinfo{journal}{J. Chem. Phys.}
  \textbf{\bibinfo{volume}{109}}, \bibinfo{pages}{834} (\bibinfo{year}{1998}).

\bibitem[{\citenamefont{Dudowicz et~al.}(1999)\citenamefont{Dudowicz, Freed,
  and Douglas}}]{douglas}
\bibinfo{author}{\bibfnamefont{J.}~\bibnamefont{Dudowicz}},
  \bibinfo{author}{\bibfnamefont{K.~F.} \bibnamefont{Freed}}, \bibnamefont{and}
  \bibinfo{author}{\bibfnamefont{J.~F.} \bibnamefont{Douglas}},
  \bibinfo{journal}{J. Chem. Phys.} \textbf{\bibinfo{volume}{111}},
  \bibinfo{pages}{7116} (\bibinfo{year}{1999}).

\bibitem[{\citenamefont{Dudowicz et~al.}(2003)\citenamefont{Dudowicz, Freed,
  and Douglas}}]{Dudo_03}
\bibinfo{author}{\bibfnamefont{J.}~\bibnamefont{Dudowicz}},
  \bibinfo{author}{\bibfnamefont{K.~F.} \bibnamefont{Freed}}, \bibnamefont{and}
  \bibinfo{author}{\bibfnamefont{J.~F.} \bibnamefont{Douglas}},
  \bibinfo{journal}{J. Chem. Phys.} \textbf{\bibinfo{volume}{119}},
  \bibinfo{pages}{12645} (\bibinfo{year}{2003}).

\bibitem[{\citenamefont{Coniglio and Klein}(1980)}]{coniglio}
\bibinfo{author}{\bibfnamefont{A.}~\bibnamefont{Coniglio}} \bibnamefont{and}
  \bibinfo{author}{\bibfnamefont{W.}~\bibnamefont{Klein}}, \bibinfo{journal}{J.
  Phys. A} \textbf{\bibinfo{volume}{13}}, \bibinfo{pages}{2775}
  (\bibinfo{year}{1980}).

\bibitem[{\citenamefont{{Zaccarelli} et~al.}(2007)\citenamefont{{Zaccarelli},
  {Sciortino}, and {Tartaglia}}}]{prossimo}
\bibinfo{author}{\bibfnamefont{E.}~\bibnamefont{{Zaccarelli}}},
  \bibinfo{author}{\bibfnamefont{F.}~\bibnamefont{{Sciortino}}},
  \bibnamefont{and}
  \bibinfo{author}{\bibfnamefont{P.}~\bibnamefont{{Tartaglia}}},
  \bibinfo{journal}{J. Chem. Phys.} \textbf{\bibinfo{volume}{127}},
  \bibinfo{pages}{174501} (\bibinfo{year}{2007}).

\bibitem[{\citenamefont{Panagiotopoulos}(1987)}]{GEMC}
\bibinfo{author}{\bibfnamefont{A.~Z.} \bibnamefont{Panagiotopoulos}},
  \bibinfo{journal}{Mol. Phys.} \textbf{\bibinfo{volume}{61}},
  \bibinfo{pages}{813} (\bibinfo{year}{1987}).

\bibitem[{\citenamefont{Ferrenberg and Swendsen}(1988)}]{histrew}
\bibinfo{author}{\bibfnamefont{A.~M.} \bibnamefont{Ferrenberg}}
  \bibnamefont{and} \bibinfo{author}{\bibfnamefont{R.~H.}
  \bibnamefont{Swendsen}}, \bibinfo{journal}{Phys. Rev. Lett.}
  \textbf{\bibinfo{volume}{61}}, \bibinfo{pages}{2635} (\bibinfo{year}{1988}).

\bibitem[{\citenamefont{Wilding}(1996)}]{Wilding_96}
\bibinfo{author}{\bibfnamefont{N.~B.} \bibnamefont{Wilding}},
  \bibinfo{journal}{J. Phys.: Condens. Matter} \textbf{\bibinfo{volume}{{\bf
  9}}}, \bibinfo{pages}{585} (\bibinfo{year}{1996}).

\bibitem[{\citenamefont{Kindt}(2002)}]{Kindt}
\bibinfo{author}{\bibfnamefont{J.~T.} \bibnamefont{Kindt}},
  \bibinfo{journal}{J. Phys. Chem. B} \textbf{\bibinfo{volume}{106}},
  \bibinfo{pages}{8223} (\bibinfo{year}{2002}).

\bibitem[{\citenamefont{Foffi and Sciortino}(2007)}]{FoffiKern}
\bibinfo{author}{\bibfnamefont{G.}~\bibnamefont{Foffi}} \bibnamefont{and}
  \bibinfo{author}{\bibfnamefont{F.}~\bibnamefont{Sciortino}},
  \bibinfo{journal}{J. Phys. Chem. B} \textbf{\bibinfo{volume}{111}},
  \bibinfo{pages}{9702} (\bibinfo{year}{2007}).

\bibitem[{\citenamefont{Kern and D.Frenkel}(2003)}]{Kern_03}
\bibinfo{author}{\bibfnamefont{N.}~\bibnamefont{Kern}} \bibnamefont{and}
  \bibinfo{author}{\bibnamefont{D.Frenkel}}, \bibinfo{journal}{J. Chem. Phys.}
  \textbf{\bibinfo{volume}{{\bf 118}}}, \bibinfo{pages}{9882}
  (\bibinfo{year}{2003}).

\bibitem[{\citenamefont{Rubinstein and Colby}(2003)}]{colby}
\bibinfo{author}{\bibfnamefont{M.}~\bibnamefont{Rubinstein}} \bibnamefont{and}
  \bibinfo{author}{\bibfnamefont{R.~H.} \bibnamefont{Colby}},
  \emph{\bibinfo{title}{Polymer Physics}} (\bibinfo{publisher}{Oxford
  University Press Inc., New York}, \bibinfo{year}{2003}).

\bibitem[{\citenamefont{{Zilman} et~al.}(2003)\citenamefont{{Zilman}, {Tlusty},
  and {Safran}}}]{safran}
\bibinfo{author}{\bibfnamefont{A.}~\bibnamefont{{Zilman}}},
  \bibinfo{author}{\bibfnamefont{T.}~\bibnamefont{{Tlusty}}}, \bibnamefont{and}
  \bibinfo{author}{\bibfnamefont{S.~A.} \bibnamefont{{Safran}}},
  \bibinfo{journal}{J. Phys.: Condens. Matter} \textbf{\bibinfo{volume}{15}},
  \bibinfo{pages}{57} (\bibinfo{year}{2003}).

\bibitem[{\citenamefont{{Stukalin} and {Freed}}(2006)}]{karlcn}
\bibinfo{author}{\bibfnamefont{E.~B.} \bibnamefont{{Stukalin}}}
  \bibnamefont{and} \bibinfo{author}{\bibfnamefont{K.~F.}
  \bibnamefont{{Freed}}}, \bibinfo{journal}{J. Chem. Phys.}
  \textbf{\bibinfo{volume}{125}}, \bibinfo{pages}{4905} (\bibinfo{year}{2006}).

\bibitem[{\citenamefont{Workum and Douglas}(2006)}]{Workum_06}
\bibinfo{author}{\bibfnamefont{V.}~\bibnamefont{Workum}} \bibnamefont{and}
  \bibinfo{author}{\bibfnamefont{J.~F.} \bibnamefont{Douglas}},
  \bibinfo{journal}{Phys. Rev. E} \textbf{\bibinfo{volume}{73}},
  \bibinfo{pages}{031502} (\bibinfo{year}{2006}).

\bibitem[{\citenamefont{{Marques} and {Cates}}(1991)}]{cates1}
\bibinfo{author}{\bibfnamefont{C.~M.} \bibnamefont{{Marques}}}
  \bibnamefont{and} \bibinfo{author}{\bibfnamefont{M.~E.}
  \bibnamefont{{Cates}}}, \bibinfo{journal}{Journal de Physique II}
  \textbf{\bibinfo{volume}{1}}, \bibinfo{pages}{489} (\bibinfo{year}{1991}).

\bibitem[{\citenamefont{{Rouault} and {Milchev}}(1996)}]{milchevdyn}
\bibinfo{author}{\bibfnamefont{Y.}~\bibnamefont{{Rouault}}} \bibnamefont{and}
  \bibinfo{author}{\bibfnamefont{A.}~\bibnamefont{{Milchev}}},
  \bibinfo{journal}{Europhysics Letters} \textbf{\bibinfo{volume}{33}},
  \bibinfo{pages}{341} (\bibinfo{year}{1996}).

\bibitem[{\citenamefont{Tartaglia}(2007)}]{nuovotartaglia}
\bibinfo{author}{\bibfnamefont{P.}~\bibnamefont{Tartaglia}},
  \bibinfo{journal}{AIP Conference Proceedings of the 5th International
  Workshop on Complex Systems, Sendai, Japan, in press}
  (\bibinfo{year}{2007}).

\bibitem[{\citenamefont{{Sastry}}(2000)}]{sastryprl}
\bibinfo{author}{\bibfnamefont{S.}~\bibnamefont{{Sastry}}},
  \bibinfo{journal}{Phys. Rev. Lett.} \textbf{\bibinfo{volume}{85}},
  \bibinfo{pages}{590} (\bibinfo{year}{2000}).

\bibitem[{\citenamefont{{De~Michele}
  et~al.}(2006{\natexlab{b}})\citenamefont{{De~Michele}, Tartaglia, and
  Sciortino}}]{cristianosilica}
\bibinfo{author}{\bibfnamefont{C.}~\bibnamefont{{De~Michele}}},
  \bibinfo{author}{\bibfnamefont{P.}~\bibnamefont{Tartaglia}},
  \bibnamefont{and}
  \bibinfo{author}{\bibfnamefont{F.}~\bibnamefont{Sciortino}},
  \bibinfo{journal}{J. Chem. Phys.} \textbf{\bibinfo{volume}{125}},
  \bibinfo{pages}{204710} (\bibinfo{year}{2006}{\natexlab{b}}).

\bibitem[{\citenamefont{Sciortino}(2007)}]{statphys}
\bibinfo{author}{\bibfnamefont{F.}~\bibnamefont{Sciortino}},
  \bibinfo{journal}{Proceedings of Stat-Phys XXIII, Eur. Phys. J. B in press}
  (\bibinfo{year}{2007}).

\bibitem[{\citenamefont{Zaccarelli et~al.}(2006)\citenamefont{Zaccarelli,
  Saika-Voivod, Buldyrev, Moreno, Tartaglia, and Sciortino}}]{Zacca2}
\bibinfo{author}{\bibfnamefont{E.}~\bibnamefont{Zaccarelli}},
  \bibinfo{author}{\bibfnamefont{I.}~\bibnamefont{Saika-Voivod}},
  \bibinfo{author}{\bibfnamefont{S.~V.} \bibnamefont{Buldyrev}},
  \bibinfo{author}{\bibfnamefont{A.~J.} \bibnamefont{Moreno}},
  \bibinfo{author}{\bibfnamefont{P.}~\bibnamefont{Tartaglia}},
  \bibnamefont{and}
  \bibinfo{author}{\bibfnamefont{F.}~\bibnamefont{Sciortino}},
  \bibinfo{journal}{J. Chem. Phys.} \textbf{\bibinfo{volume}{{\bf 124}}},
  \bibinfo{pages}{124908} (\bibinfo{year}{2006}).

\bibitem[{\citenamefont{Padoa-Schioppa
  et~al.}(1998)\citenamefont{Padoa-Schioppa, Sciortino, and Tartaglia}}]{padua}
\bibinfo{author}{\bibfnamefont{C.}~\bibnamefont{Padoa-Schioppa}},
  \bibinfo{author}{\bibfnamefont{F.}~\bibnamefont{Sciortino}},
  \bibnamefont{and}
  \bibinfo{author}{\bibfnamefont{P.}~\bibnamefont{Tartaglia}},
  \bibinfo{journal}{Phys. Rev. E} \textbf{\bibinfo{volume}{57}},
  \bibinfo{pages}{3797 } (\bibinfo{year}{1998}).

\end{thebibliography}

\clearpage

\begin{table}[h]
\begin{center}
\begin{tabular}{|c|c|c|c|c|c|c|}
\hline
$f$ &   $T_c$  &  $\rho_c$  &  $\mu_c$  &  $s$  & $L$ & $B_2^c/B_2^{HS}$ \\
\hline
\hline
3  geometric   &       0.094  &  0.141  &  -0.471  &  0.46    & 9 & -28.772  \\
4  geometric   &       0.118  &  0.273  &  -0.418  &  0.08    & 7 & -5.080  \\
5  geometric   &       0.132  &  0.351  &  -0.410  &  0       & 7 & -2.866 \\  
\hline
4 random &       0.102  &  0.208  &  -0.531  &  0.46    & 8  & -21.978 \\  
5 random &       0.118  &  0.258  &  -0.500  &  0.25    & 8  & -8.500 \\
6 random &       0.133  &  0.310  &  -0.482  &  0.22    & 7  & -4.258 \\ 
\hline 
\end{tabular}
\end{center}
\caption{Values of the relevant parameters at the critical point for the geometric~\cite{bian} $(f=3,4,5)$ and random $(f=4,5,6)$ cases: $T_c$ is the critical temperature, $\rho_c$ is the density of the critical point, $\mu_c$ is the critical chemical potential, $s$ is the field mixing parameter and $L$ indicates the largest studied box size. $B_2^c/B_2^{HS}$ is the value of the reduced second virial coefficient at the critical point.}
\label{table:total}\end{table}

\clearpage
\begin{table}[h]
\begin{center}
\begin{tabular}{|c|c|c|c|c|c|c|}
\hline
   $f$ &   $T_c$  &  $\rho_c$ & $B_2^c/B_2^{HS}$\\
   \hline
    \hline
3  theory     &       0.0925  &  0.086 & -34.378 \\
4  theory     &       0.1121  &  0.154 & -8.498 \\
5  theory     &       0.1275  &  0.212 & -4.052 \\
6  theory     &       0.1411  &  0.261 & -2.414 \\
 \hline 
\end{tabular}
\end{center}
\caption{Critical values of the temperature and density for $3 \leq f \leq 6$ evaluated through the Wertheim theory. We also report the corresponding values of the reduced second virial coefficient.}
\label{table:teo}
\end{table}

\clearpage
\begin{table}[h]
\begin{center}
\begin{tabular}{|c|c|c|c|c|c|c|}
 \hline 
\hspace{0.01cm}  $f$ \hspace{0.04cm} & \hspace{0.01cm}$p_b^c$ \hspace{0.06cm} $theory$  \hspace{0.01cm}&  \hspace{0.01cm} $p_b^c$\hspace{0.06cm} $geometric$  \hspace{0.01cm} & \hspace{0.01cm} $p_b^c$\hspace{0.06cm} $random$ \hspace{0.01cm} \\
\hline
  \hline 
3 &     0.633  &  0.728   &         \\
4 &     0.488  &  0.640   &  0.737  \\
5 &     0.417  &  0.577   &  0.615  \\
6 &     0.360  &          &  0.539  \\
  \hline 
\end{tabular}
\end{center}
\caption{Critical values of the bond probability $p_b$ for $f$ varying from $3$ to $6$. Theoretical values are obtained solving Eq.~\ref{eq:pb} at the critical point, while numerical ones are obtained as the ratio between the potential energy at the critical point and the energy of the fully bonded system.}
\label{table:pb}
\end{table}

\clearpage

\begin{figure}[h] 
\includegraphics[width=10cm, clip=true]{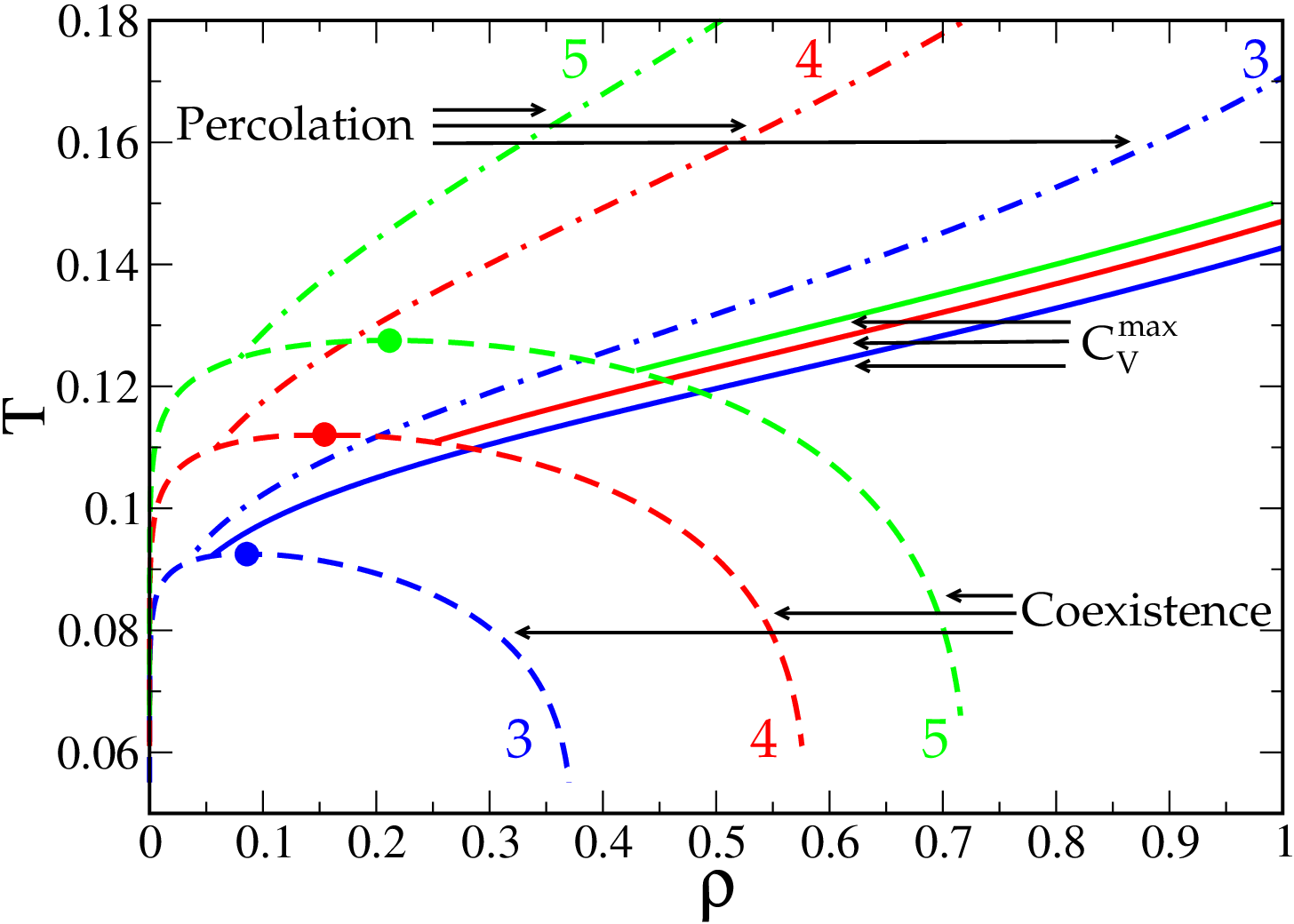}
  \caption{
}
   \label{fig:pdWtot}
\end{figure}

\clearpage

\begin{figure}[h] 
\includegraphics[width=11cm, clip=true]{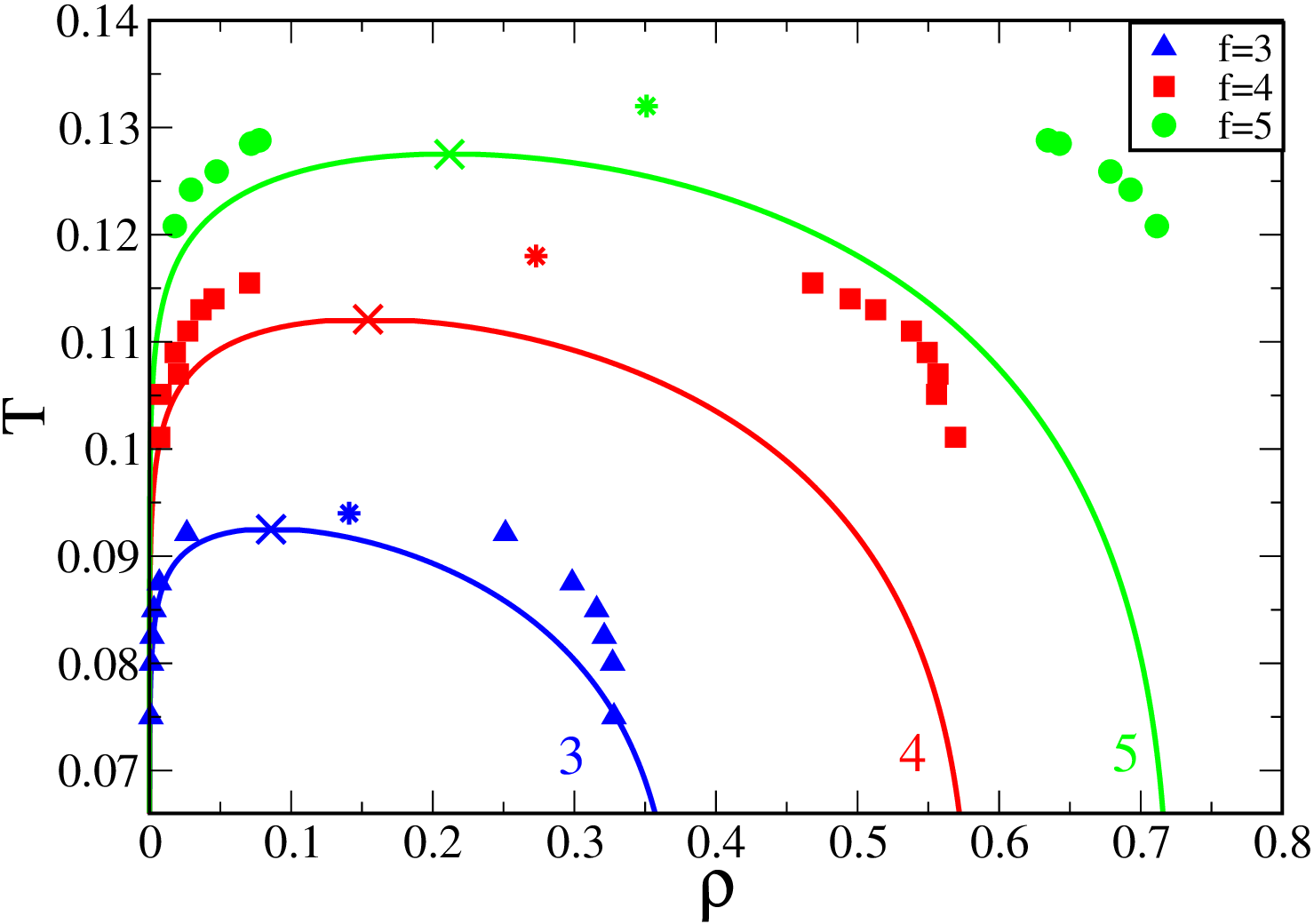}
   \caption{
}
   \label{fig:pdW}
\end{figure}

\clearpage

\begin{figure}[h] 
\includegraphics[width=10cm, clip=true]{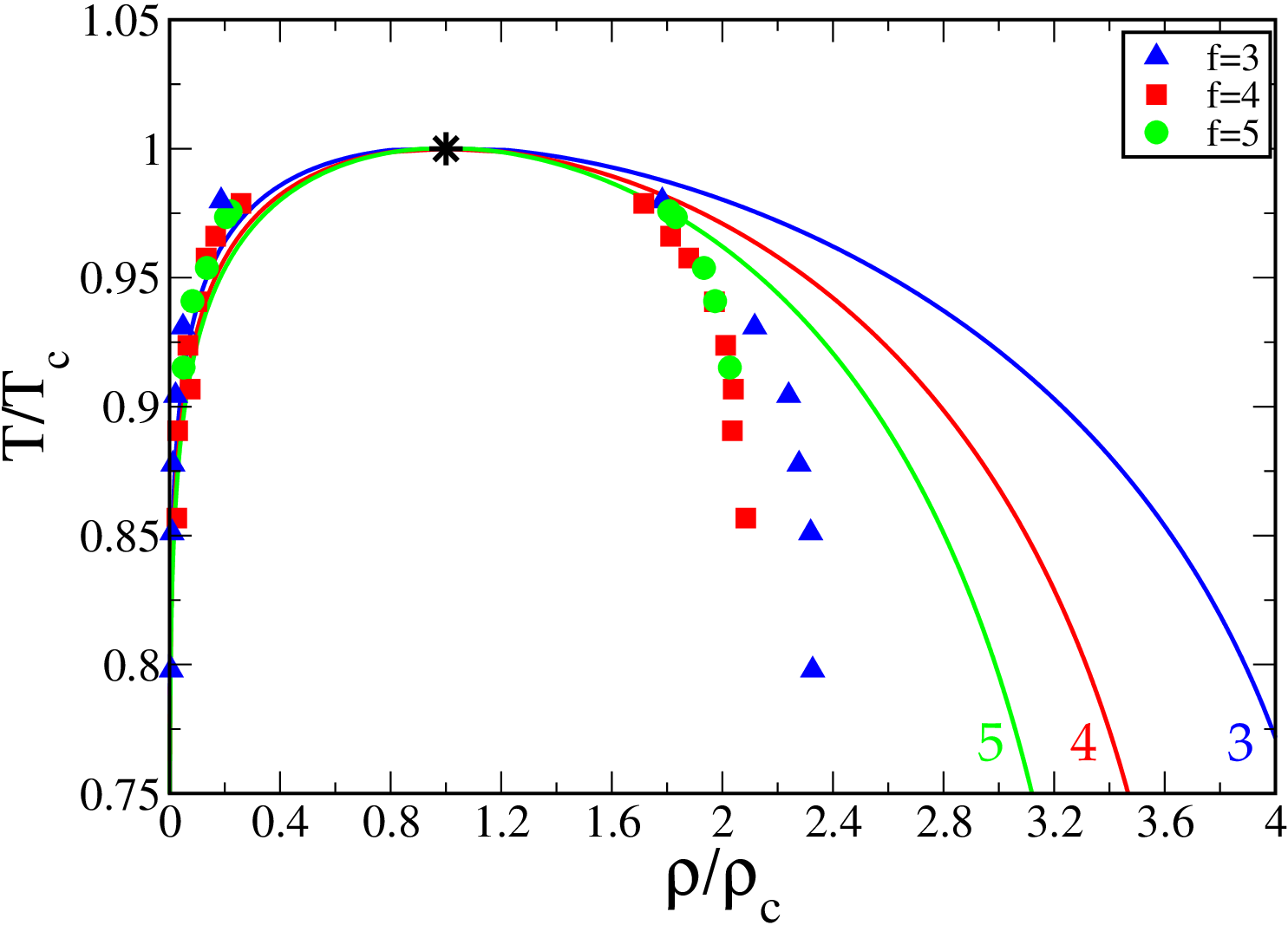}
   \caption{
}
   \label{fig:pdscaled}
\end{figure}

\clearpage

\begin{figure}[h] 
 \includegraphics[width=10cm, clip=true]{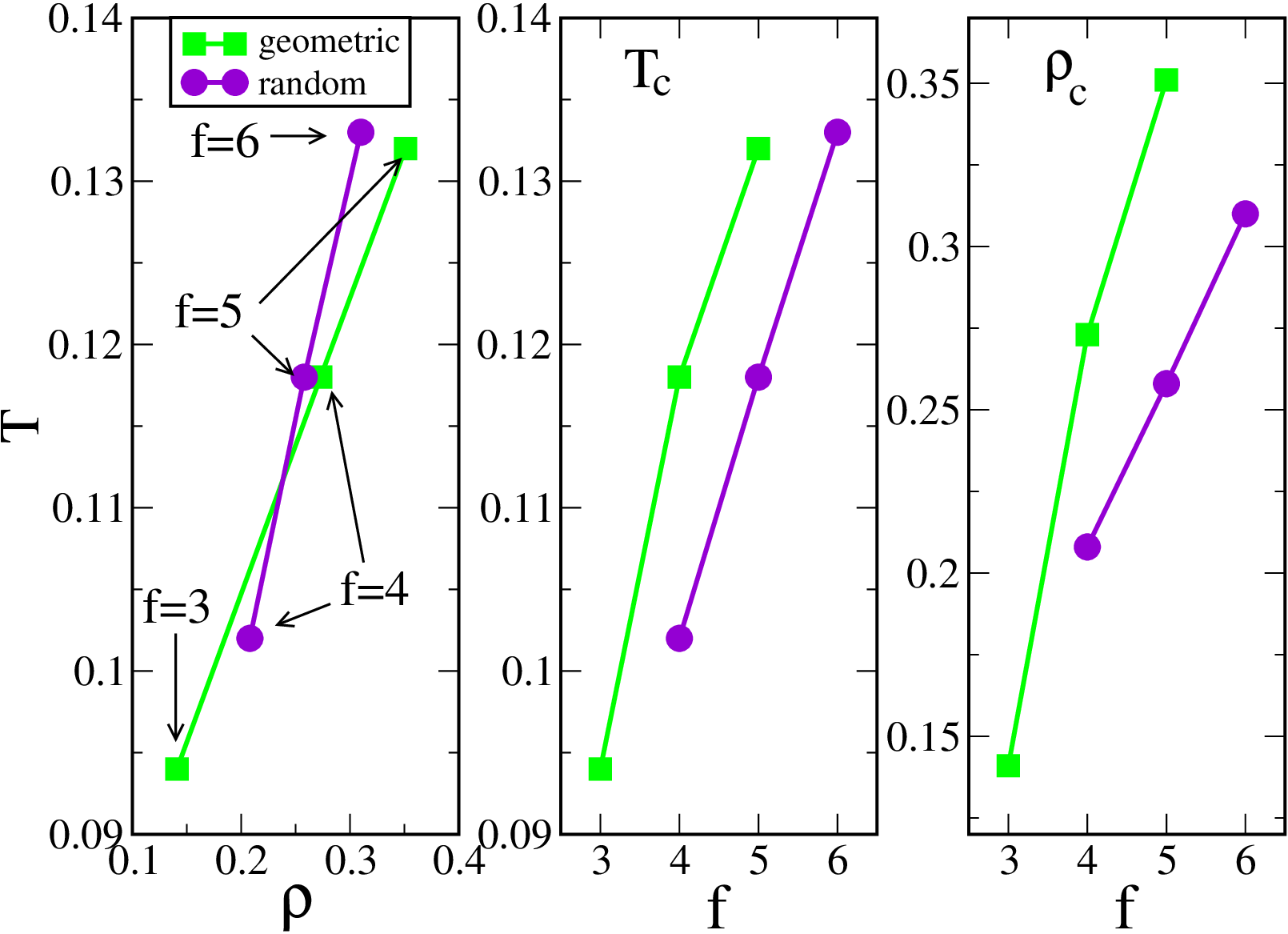}
   \caption{
}
   \label{fig:tcphic}
\end{figure}

\clearpage
\begin{figure}[h] 
 \includegraphics[width=10cm, clip=true]{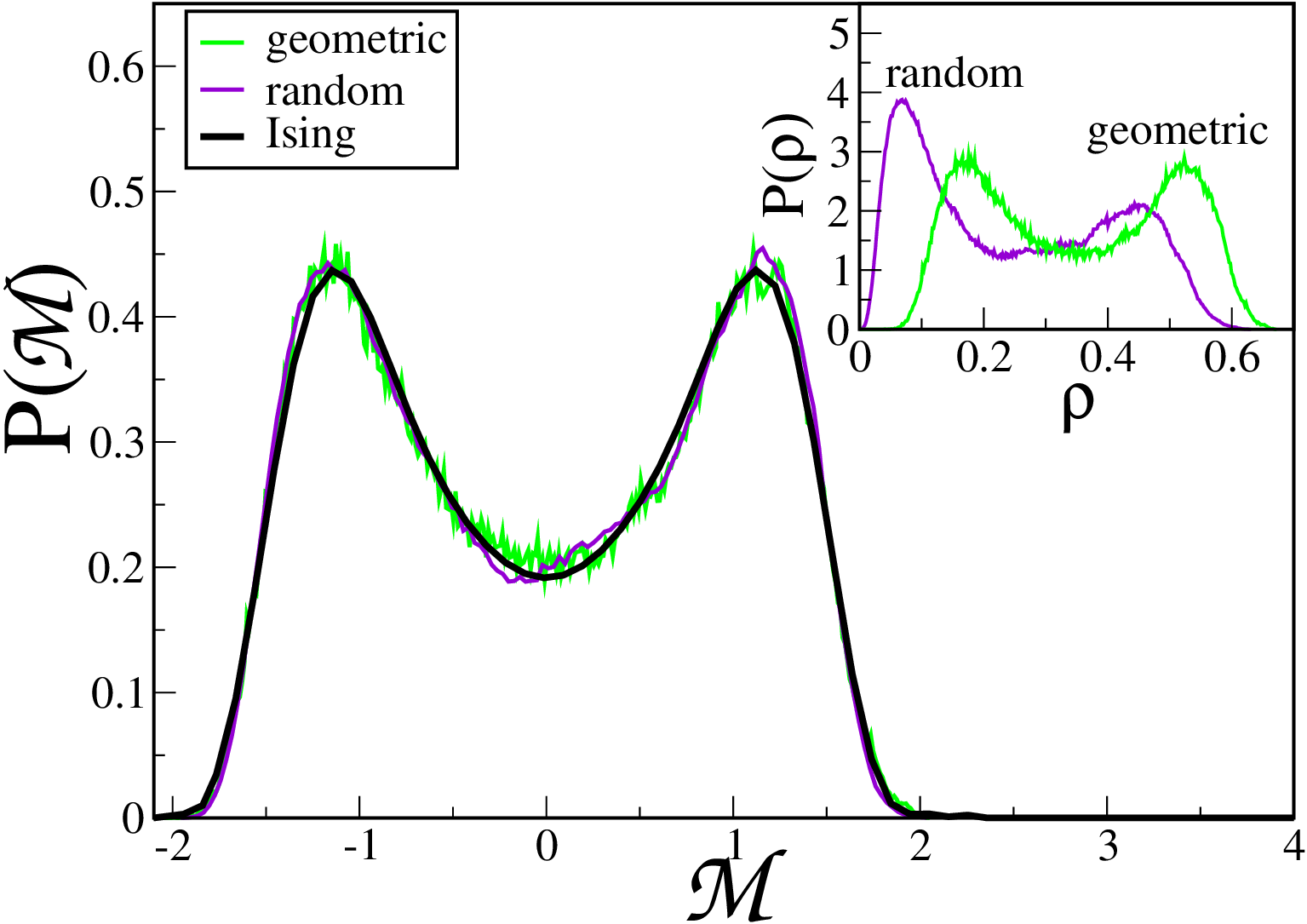}
   \caption{
}
\label{fig:rw5}
\end{figure}

\clearpage
\begin{figure}[h] 
\includegraphics[width=10cm, clip=true]{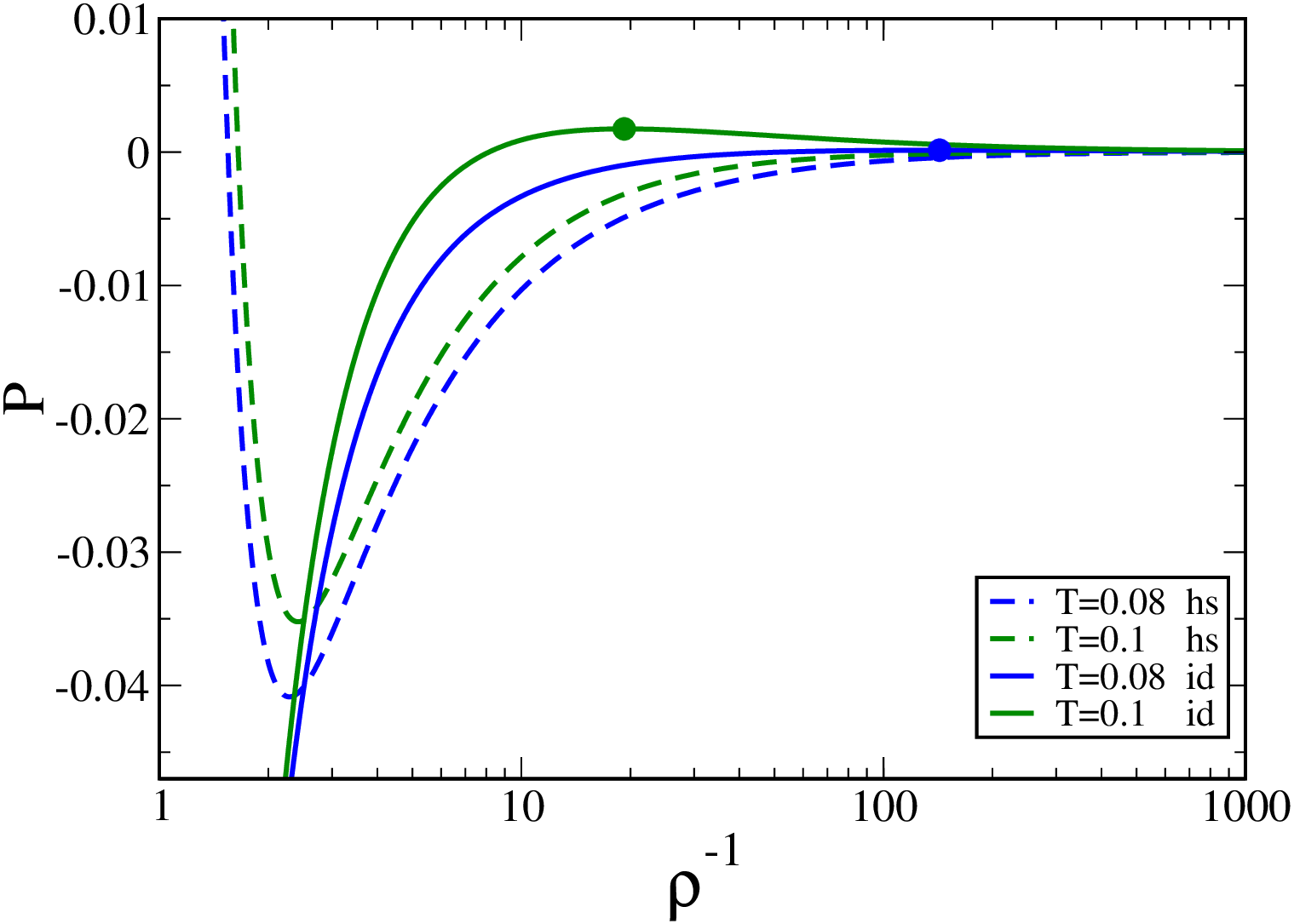}
   \caption{
}
   \label{fig:p}
\end{figure}

\clearpage

\setcounter{figure}{0}

\begin{figure}[h] 
   \caption{Theoretical predictions for the phase diagram of patchy systems on varying the  particles functionality form $f=3$ to $5$. Coexistence curves and $C_V^{max}$ lines are evaluated according to the Wertheim theory, respectively from Eq.~\ref{eq:coex} and by finding the zeroes of the temperature derivative of Eq.~\ref{eq:cv}, i.e. $(\partial C_V/\partial T)_V=0$. Percolation lines are evaluated according to the Flory-Stockmayer theory as the locus of points in the $(T,\rho)$ plane such that $p_b(T,\rho)=p_b^p$ (see Eq.~\ref{eq:perc}).}
   \label{fig:pdWtot}
\end{figure}

\begin{figure}[h] 
   \caption{Gas-liquid coexistence regions  in the $(T,\rho)$ plane on varing $f$ from $3$ to $5$. Points are GEMC numerical results for the model in which the sticky sites are geometrically arranged on the particles surface. Solid lines are Wertheim theoretical predictions for the coexistence curves obtained from Eq.~\ref{eq:coex}. The numerical (stars) and theoretical (crosses) critical points  from~\cite{bian} are reported to help visualizing the critical behavior. }
   \label{fig:pdW}
\end{figure}

\begin{figure}[h] 
   \caption{Gas-liquid coexistence curves in terms of the reduced temperature $T/T_c$ and the reduced density $\rho/\rho_c$ for systems with $f=3,4,5$ sticky sites. Points are GEMC results for the geometric arrangements of the patches while solid lines are the Wertheim theoretical predictions.}
   \label{fig:pdscaled}
\end{figure}

\begin{figure}[h] 
   \caption{Comparison  between numerical results for
   patchy particles with different number of sticky spots per particle in a geometric (squares) and random (circles) arrangement.   Panel (a) shows the location of the points in the $(T,\rho)$ plane. Panels (b) and (c) compare respectively the  $f$ dependence for $T_c$ and $\rho_c$. Data for the ordered case are reproduced from Ref.~\cite{bian}.}
   \label{fig:tcphic}
\end{figure}

\begin{figure}[h] 
   \caption{Comparison between the critical fluctuations distributions of $\mathcal{M} \sim \rho + s u$ in both the geometric and random cases  with $f=5$. The calculated  $P(\mathcal{M})$ are compared to the expected distributions (full line) for systems at the critical point belonging to the Ising universality class~\cite{Wilding_96}. The inset shows the comparison between the corresponding density fluctuations distributions $P(\rho)$ in the two cases.
}
\label{fig:rw5}
\end{figure}

\begin{figure}[h] 
   \caption{Equation of state for two different isothermes  $(T=0.08$ and $0.1)$ in the pressure vs. volume plane in the case of three functional particles. Solid lines (id) are related to the ideal system described by Eq.~\ref{eq:pvnc}. Note that in the high $T$ limit Eq.~\ref{eq:pvnc} reduces to the ideal gas equation of state ($\beta P/ \rho=1$). Points on solid lines indicate the maxima of the pressure respect to $\rho^{-1}$: the set of these points on varing $T$ provides the spinodal line of the system. Dotted lines (hs) are related to a system in which bonding is accompained by an hard-core repulsion: the bonding contribution is given by Eq.~\ref{eq:Pbond} in which the density dependence of the $g_{HS}(r)$ (see Eq.~\ref{eq:ghsr}) is considered, while the hard-sphere contribution is given by Eq.~\ref{eq:betaPhs}.}
   \label{fig:p}
\end{figure}

\end{document}